\documentclass[twocolumn,trackchanges]{aastex631}

\usepackage{amsmath}
\usepackage{enumitem}
\usepackage{xcolor}

\shorttitle{Lepto-hadronic scenario in RS Ophiuchi}
\shortauthors{De Sarkar et al.}
\begin{document}

\title{Lepto-hadronic interpretation of 2021 RS Ophiuchi nova outburst}

\author[0000-0001-6047-6746]{Agnibha De Sarkar}
\affiliation{Astronomy $\&$ Astrophysics group, Raman Research Institute \\
C. V. Raman Avenue, 5th Cross Road, Sadashivanagar, Bengaluru 560080, Karnataka, India}
\email{agnibha@rri.res.in}

\author[0000-0002-8070-5400]{Nayana A. J.}
\affiliation{Indian Institute of Astrophysics, II Block, Koramangala, Bangalore 560034, India}

\author[0000-0001-9829-7727]{Nirupam Roy}
\affiliation{Dept. of Physics, Indian Institute of Science \\
CV Raman Road, Bengaluru 560012, Karnataka, India}

\author[0000-0002-0130-2460]{Soebur Razzaque}
\affiliation{Centre for Astro-Particle Physics (CAPP) and Department of Physics, University of Johannesburg, PO Box 524, Auckland Park
2006, South Africa}
\affiliation{ Department of Physics, The George Washington University, Washington, DC 20052, USA}
\affiliation{National Institute for Theoretical and Computational Sciences (NITheCS), South Africa}

\author[0000-0003-3533-7183]{G. C. Anupama}
\affiliation{Indian Institute of Astrophysics, II Block, Koramangala, Bangalore 560034, India}


\begin{abstract}

Very high energy (VHE; 100 GeV $<$ E $\leq$ 100 TeV) and high energy (HE; 100 MeV $<$ E $\leq$ 100 GeV) gamma-rays were observed from the symbiotic recurrent nova RS Ophiuchi (RS Oph) during its outburst in August 2021, by various observatories such as High Energy Stereoscopic System (H.E.S.S.), Major Atmospheric Gamma Imaging Cherenkov (MAGIC), and {\it Fermi}-Large Area Telescope (LAT). The models explored so far tend to favor a hadronic scenario of particle acceleration over an alternative leptonic scenario. This paper explores a time-dependent lepto-hadronic scenario to explain the emission from the RS Oph source region. We have used simultaneous low frequency radio data observed by various observatories, along with the data provided by H.E.S.S., MAGIC, and \textit{Fermi}-LAT, to explain the multi-wavelength (MWL) spectral energy distributions (SEDs) corresponding to 4 days after the outburst. Our results show that a lepto-hadronic interpretation of the source not only explains the observed HE-VHE gamma-ray data but the corresponding model synchrotron component is also consistent with the first 4 days of low radio frequency data, indicating the presence of non-thermal radio emission at the initial stage of nova outburst. We have also calculated the expected neutrino flux from the source region and discussed the possibility of detecting neutrinos.  

\end{abstract}

\keywords{Recurrent novae (1366) --- Symbiotic novae (1675) --- Gamma-ray sources (633) --- Radio sources (1358)}

\section{Introduction} \label{sec1}

Nova outburst happens in a binary star system comprising a white dwarf (WD) as the compact object and a companion star. The material from the companion star gets accreted on the WD surface. When enough layers of material have accumulated, it eventually causes a thermonuclear runaway explosion on the surface of the WD. The subsequent eruption ejects the bulk of the accreted material at a few thousand km/s and brightens up the WD up to $\sim$ 10$^{4-5}$ L$_{\odot}$ \citep{gomez98, hellier01, warner03, knigge11}. While the companion in nova systems is, in general, a low mass, main sequence late-type star \citep{bode08, chomiuk21}, in some cases the companion is a Red Giant (RG) (or sub-giant, in general, an evolved star). Such systems are classified as a symbiotic nova \citep{shore11, shore12, miko12}. Since the ejected material from the WD surface produces shock in the ambient medium, the nova outburst phenomenon provides the extreme conditions needed to accelerate particles. Shock can occur when slow-moving ejecta collides with faster-moving ejecta (internal shock), as observed in classical novae. Alternatively, fast-moving outflow can collide with pre-existing dense wind of the RG star and produce a shock (external shock), typically considered to be happening in symbiotic novae. Since nova outburst is a hotbed for particle acceleration, gamma-ray emission is also expected to be observed from such transient phenomenon.

Although novae have long been observed in the optical wavelengths, during the last decade, over a dozen novae have been detected at GeV gamma-ray energies by \textit{Fermi}-LAT \citep{ackermann14, cheung16, franck18, gordon21, chomiuk21}, starting with the detection of GeV gamma-rays from the nova eruption of V407 Cygni \citep{abdo10} in 2010. The spectra of these sources were usually well described by a simple power law with a spectral index close to $-2$ and an exponential cutoff of a few GeV. However, the recent gamma-ray detection of RS Ophiuchi (RS Oph) nova outburst on August 2021 by MAGIC and H.E.S.S. has provided conclusive evidence that the particles during nova outburst can be accelerated to TeV energies \citep{magic22, hess22}.  

RS Oph is a symbiotic recurrent nova that shows nova outbursts every 15-20 years \citep{dob94}. The binary system comprises a massive WD (1.2 - 1.4 M$_{\odot}$) and an RG star usually identified as M0-2 III \citep{anupama99, dob94, barry08, brandi09}. The distance of RS Oph is a matter of intense debate. The distance of the source has been posited to be D = 1.4 kpc \citep{barry08, hess22}, as well as D = 2.45 kpc \citep{magic22}. In this particular work, we have considered the distance of the source to be 2.45 kpc, following \cite{magic22}. The binary separation of the components is approximately 1.48 Astronomical Units (AU) \citep{booth16, fermi22}. GeV-TeV emission from symbiotic novae, especially RS Oph, was previously predicted in \cite{tatischeff07}, \cite{tatischeff08}, but it had not been detected earlier. On 8th August 2021, the outburst of RS Oph was detected in optical observations at a visual magnitude of 4.5 \citep{kafka}. Subsequently, MAGIC and H.E.S.S. reported TeV gamma-rays from the source region up to 4-5 days after the outburst \citep{magic22, hess22}. \textit{Fermi}-LAT data analysis also provided GeV gamma-ray detection from the source region \citep{fermi22}. The detection of HE-VHE gamma-rays from the outburst indicates particle acceleration in the source region and confirms that novae can be TeVatrons. The production of gamma-rays can be explained using accelerated protons colliding with the downstream gas (hadronic $\pi^0$ decay model) or energetic electrons scattering low energy photons in the nova photosphere (leptonic inverse Compton (IC) model). The hadronic nature of gamma-ray emission has been generally preferred in previous studies of RS Oph \citep{magic22, hess22}, although a leptonic scenario can not be entirely ruled out. \cite{hess22} have explained the HE-VHE gamma-ray emission using a single shock, single particle population model. On the contrary, \cite{diesing22} have explained the HE-VHE gamma-ray emission using multiple shocks, single particle population scenario. Note that none of these models considered lower energy data points (e.g. radio). A purely leptonic scenario has been neglected in these models, citing that leptons will lose energy very efficiently and will not be able to produce the observed HE-VHE gamma-ray emission.

This work proposes an alternative approach to explain the HE-VHE gamma-ray emission with a single shock, multiple particle population (lepto-hadronic) scenario. This scenario was explored previously in \cite{sitarek12, martin13, magic15} for the cases of novae V407 Cygni and V339 Del. However, the lack of any significant VHE gamma-ray data rendered this scenario inconclusive for the cases explored at that time. With the VHE gamma-ray data from MAGIC and H.E.S.S., and HE gamma-ray data from \textit{Fermi}-LAT, we try to explore the feasibility of a lepto-hadronic scenario of particle acceleration in RS Oph nova eruption. Considering that it is unlikely that the entire HE-VHE gamma-ray emission is produced from a purely leptonic scenario, we propose that the HE gamma-rays observed from the source region are produced by IC cooling of accelerated leptons, whereas VHE gamma-rays are hadronic in nature. In addition, we have also used the observed low frequency radio data to create multi-wavelength (MWL) spectral energy distribution (SED) for the first 4 days after the outburst. These MWL SEDs were explained by solving a time-dependent, diffusion-loss equation iteratively, using the open-source code \texttt{GAMERA} \citep{hahn}. The combined lepto-hadronic model explored in this paper satisfactorily explains the MWL SEDs in the HE-VHE gamma-ray range for all 4 days. Additionally, the corresponding model synchrotron emission is consistent with first 4 days of radio data as well, which essentially indicates that the total low frequency radio emission from the region was non-thermal synchrotron dominated during the first few days after the outburst, at the very least. The presence of synchrotron emission during the first 4 days after the outburst confirms that presence of accelerated electrons near the forward shock. Well-sampled, early time MWL analyses of future nova outbursts are necessary to confirm the actual particle acceleration scenario(s) in such sources. Moreover, we have calculated the total neutrino flux expected from the source region in our model and discussed the probability of neutrino detection from RS Oph by the next-generation neutrino detectors.

We explain the simple model explored in this work in Section \ref{sec2}, present our results in Section \ref{sec3}, discuss the obtained results in Section \ref{sec4}, and finally, conclude in Section \ref{sec5}.

\section{The model} \label{sec2}

A symbiotic recurrent nova such as RS Oph is a complex hydrodynamic process in which shock results from the interaction between high-velocity ejecta and surrounding circumbinary medium and behaves as a small-scale supernova system, but with a shorter timescale of a few weeks to months. The interaction between the shock and the stellar wind in the circumbinary medium produces both forward and reverse shocks, separated by a contact discontinuity. We only consider the forward shock for explaining the gamma-ray emission, similar to \cite{magic22, hess22, zheng22}. During the early evolution of the nova, the ejecta is in the free expansion phase, and the shock moves at a constant velocity. The corresponding wind density is shaped as $\rho$ $\propto$ ($r^2$ + $a^2$ - $2arcos \theta$)$^{-1}$, where $a$ is the semi-major axis of the binary system and $r$ is the distance from the WD. So the wind structure at a small radius is aspherical, centered at the WD. However, at $r >> a$ the wind density structure is close to a spherical structure with $\rho$ $\propto$ $r^{-2}$. In this work, we have considered the spherical approximation of the wind structure for simplicity.

When enough material gets swept up by the expanding shock, the free expansion stops, and the shock decelerates. Subsequently, the shock enters the energy-conserving, adiabatic, Sedov-Taylor phase. During this phase, the shock radius and velocity evolve as r$_{sh}$ $\propto$ t$^{2/3}$ and v$_{sh}$ $\propto$ t$^{-1/3}$ respectively, for the stellar wind case \citep{zheng22}. Subsequently, the deceleration rate increases further when the shock goes into momentum conserving, radiative phase, where radiative cooling starts to dominate. During this phase, the shock radius and velocity evolve as r$_{sh}$ $\propto$ t$^{1/2}$ and v$_{sh}$ $\propto$ t$^{-1/2}$ respectively, for the stellar wind case \citep{zheng22}. We are only interested in the first 4 days after the outburst, during which the shock remains in the adiabatic phase \citep{hess22, zheng22}. In this work, we assume a day 1 shock radius (r$_{sh, 1}$) and velocity (v$_{sh, 1}$) of 4 $\times$ 10$^{13}$ cm and 4500 km/s, respectively following \cite{magic22}. Apart from that, we assume ejected mass (M$_{ej}$) = 10$^{-6}$M$_{\odot}$, mass loss rate of the RG star (\.{M$_{RG}$}) = 5 $\times$ 10$^{-7}$ M$_{\odot}$ yr$^{-1}$, and velocity of the RG wind (v$_{RG}$) = 10 km/s to be fixed in this work \citep{magic22}. For an adiabatic shock, the shock radius and velocity evolve in time as,

\begin{equation}\label{eq1}
    r_{sh}(t) = r_{sh, 1}\left( \frac{t}{1\:d} \right)^{\frac{2}{3}}
\end{equation}

\begin{equation}\label{eq2}
    v_{sh}(t) = v_{sh, 1}\left( \frac{t}{1\:d} \right)^{-\frac{1}{3}}
\end{equation}

Note that these sets of equations will not be valid at later stages of the evolution, and one has to consider radiative shock conditions.

Particles get accelerated at the adiabatic shock through the diffusive shock acceleration (DSA) mechanism. The accelerated particles, i.e., protons and electrons, then interact with the ambient medium to produce observable gamma-rays. The proton population will interact with the ambient matter, either with nova ejecta or matter in the RG wind. Both ejecta matter density and RG wind density decrease as the nova shock progresses. The matter number density of the nova ejecta can be estimated as \citep{magic22},

\begin{equation}\label{eq3}
\begin{split}
    n_{ej} &= \frac{M_{ej}}{4 \pi h r_{sh}^3 m_p}\\
           &= 1.62 \times 10^{10} \left( \frac{v_{sh}}{v_{sh, 1}} \right)^{-3} \left( \frac{r_{sh}}{r_{sh, 1}} \right)^{-\frac{9}{2}} cm^{-3}
\end{split}           
\end{equation}
\\
where it has been assumed that the ejecta has concentrated at a distance of $r_{sh}$ in a layer of thickness $hr_{sh}$ with $h = 0.1$. On the other hand, the number density of the material contained in RG wind can be given by \citep{magic22},

\begin{equation}\label{eq4}
\begin{split}
    n_{RG} &= \frac{\dot{M}_{RG}}{4 \pi r_{sh}^2 v_{RG} m_p}\\
           &= 9.9 \times 10^8 \left( \frac{v_{sh}}{v_{sh, 1}} \right)^{-2} \left( \frac{r_{sh}}{r_{sh, 1}} \right)^{-3} cm^{-3}
\end{split}
\end{equation}
\\
Since we only consider the particles accelerated in the forward shock, which propagates in the RG wind, $n_{RG}$ has been used as the target proton density for hadronic \textit{p-p} interaction in this work.

Similar to protons, electrons also get accelerated in the shock. In a leptonic scenario, the observed gamma-rays can originate from IC cooling and bremsstrahlung radiation on the ambient matter. The accelerated electrons interact with the thermal, soft photons of the nova photosphere and produce gamma-rays through IC cooling. The temperature of the soft photons is considered to be $T_{ph}$ = 8460 K \citep{magic22}. Assuming a nova photosphere radius ($r_{ph}$) of 200 $R_{\odot}$, we consider the soft photon energy density to be \citep{magic22},

\begin{equation}\label{eq5}
    u_{ph} = 1.26 \left( \frac{v_{sh}}{v_{sh, 1}} \right)^{-2} \left( \frac{r_{sh}}{r_{sh, 1}} \right)^{-3} erg\:cm^{-3}
\end{equation}
\\
We have also taken into account IC emission from leptons interacting with Cosmic Microwave Background (CMB), characterized by T$_{CMB}$ = 2.7 K and u$_{CMB}$ = 0.25 eV cm$^{-3}$. Apart from interacting with soft photons, the electrons also interact with the matter in RG wind and produce gamma-rays through Bremsstrahlung radiation. We have included this component in the model as well. Note that the abovementioned equations may be crude assumptions as the parameters related to transient sources such as nova outbursts are poorly constrained. Nevertheless, we assume these conditions to remain congruent with previous literature and to investigate the conditions for which a lepto-hadronic scenario is viable in the limit of previously explored studies.  

Since we are considering electron acceleration in nova shock, it is reasonable to expect radio synchrotron emission to be present in the source region. The synchrotron emission depends on the magnetic field present in the ambient medium. A typical scenario of calculating the magnetic field strength and evolution in a nova system is not well understood. Adopting a similar prescription discussed in \cite{chomiuk12}, we get the magnetic field near the shock assuming an equipartition with the thermal energy density of the RG wind upstream of the shock,

\begin{equation}\label{eq6}
    B = \sqrt{32 \pi n_{RG} k_B T_{RG}}
\end{equation}

where, $k_{B}$ is the Boltzmann constant, and T$_{RG}$ is the temperature of the RG wind ($\sim$ 10$^3$ K). We have further assumed that the condition pertain to the wind before it was heated and subsequently ionized by the nova outburst \citep{chomiuk12}. Similar treatment was also considered in \cite{martin13}. We use equation \ref{eq6} to calculate the model synchrotron component. Note that there is an alternate way of calculating the magnetic field by assuming an equipartition with the relativistic electrons, a simplification commonly used to interpret the radio emission from supernovae \citep{chomiuk12}. However, we do not follow that approach in this work, and calculate the magnetic field near the shock following the same concept as \cite{martin13}. 

Considering all of these relations discussed above, we have solved the time-dependent, diffusion-loss (TDDL) equation iteratively, using \texttt{GAMERA} \citep{hahn}. The TDDL equation is given by,
\begin{equation}\label{eq7}
    \frac{\partial N_{e/p}}{\partial t} = Q_{e/p} (E_{e/p}, t) - \frac{\partial (b_{e/p} N_{e/p})}{\partial E_{e/p}} - \frac{N_{e/p}}{\tau^{esc}_{e/p}}
\end{equation}
where Q$_{e/p}$(E$_{e/p}$, t) signifies the injection spectrum of the considered parent electron and proton populations, b$_{e/p}$ = b(E$_{e/p}$, t) is the energy loss rates of these parent particles, $\tau^{esc}_{e/p}$ is the escape time scale and N$_{e/p}$ is the resulting particle spectrum of the system at a given time t. While solving the TDDL equation given in equation \ref{eq7}, we have considered IC, synchrotron, and bremsstrahlung cooling for electrons \citep{blumenthal, ghisellini, baring}, as well as hadronic \textit{p-p} interaction for protons \citep{kafe}. The losses due to adiabatic expansion of the source with time have also been considered. We have considered the effect of particle escape in the form of escape timescale, given by $\tau^{esc}_{e/p}$ = 2r$_{sh}$/c, where c is the velocity of light. Moreover, the full Klein-Nishina (KN) cross section was considered for the IC mechanism incorporated in the TDDL equation. Further, we assume a power law with the exponential cutoff as the injection spectrum for both electron and proton populations. The injection spectra for electrons and protons are given by,

\begin{equation}\label{eq8}
    \frac{\partial N_e}{\partial E_e} \propto E_e^{-\alpha_e} exp\left(-\frac{E_e}{E_e^{cut}} \right)
\end{equation}

and,

\begin{equation}\label{eq9}
    \frac{\partial N_p}{\partial E_p} \propto E_p^{-\alpha_p} exp\left(-\frac{E_p}{E_p^{cut}} \right)
\end{equation}

where $\alpha_e$ and $\alpha_p$ are spectral indices, and E$^{cut}_e$ and E$^{cut}_p$ are cutoff energies of the accelerated electron and proton populations, respectively.  The normalization factors (in erg$^{-1}$s$^{-1}$) of equations \ref{eq8} and \ref{eq9} are calculated through the relations,

\begin{equation}\label{eq10}
    L_e = \int E_e \frac{\partial N_e}{\partial E_e} dE_e
\end{equation}

and, 

\begin{equation}\label{eq11}
    L_p = \int E_p \frac{\partial N_p}{\partial E_p} dE_p
\end{equation}

where, L$_e$ and L$_p$ are luminosities of the electron and proton populations, respectively. The power law spectral index of the proton population $\alpha_p$ was fixed at the value of 2.2, following \cite{hess22}, whereas a somewhat harder power law spectral index of $\alpha_e$ = 1.5 was considered for the electron population, following \cite{sitarek12}. The minimum energy of the proton (electron) population has been fixed at the rest mass energy of proton (electron), i.e., E$^{min}_p$ $\sim$ 1 GeV (E$^{min}_e$ $\sim$ 0.511 MeV).

To estimate the maximum energy that the electron population can attain, we first note that the acceleration rate of the electrons is parametrized by $\dot{P}_{acc}$ = $\xi$cE/R$_L$, where R$_L$ is the Larmor radius of the particle with energy E in perpendicular magnetic field B in Gauss, the energy of electron E$_e$ is in GeV, $\xi$ = 10$^{-4}\xi_{-4}$ is the acceleration parameter and c is the speed of light. The acceleration time scale can be written as \citep{sitarek12},

\begin{equation}\label{eq12}
    \tau^e_{acc} = E_e/\dot{P}_{acc} \approx 1E_e/\xi_{-4}B\:s
\end{equation}

Also, the cooling timescales for synchrotron and IC processes (in the Thompson regime) for the electrons can be estimated as \citep{bednarek11, magic15},

\begin{equation}
    \tau_{syn} = \frac{E_e m_e^2}{\frac{4}{3} c \sigma_T u_{B} E_e^2} \approx \frac{3.7 \times 10^5}{B^2 E_e} s
\end{equation}

and,

\begin{equation}\label{eq12a}
    \tau_{IC/T} = \frac{E_e m_e^2}{\frac{4}{3} c \sigma_T u_{ph} E_e^2} \approx \frac{170}{E_e r_{ph}^2}\left[\frac{T_{ph, 4}^4}{r_{sh}^2}\right]^{-1} s
\end{equation}

where, $\sigma_T$ is the Thomson cross section, u$_B$ and u$_{ph}$ are the energy densities of the magnetic and the radiation fields, m$_e$ is the electron mass, and T$_{ph, 4}$ = T$_{ph}$/(10$^4$ K). Now, by comparing the acceleration timescale (equation \ref{eq12}) with the IC cooling time scale in the Thomson regime (equation \ref{eq12a}), we get the maximum energies that the electrons can attain as \citep{sitarek12, magic15, bednarek11},

\begin{equation}\label{eq13}
    E^{max}_e \approx 13(\xi_{-4}B)^{1/2}\frac{r_{sh}}{T_{ph, 4}^2 r_{ph}}\: GeV
\end{equation}

The acceleration parameter expected from a second-order Fermi acceleration on the nova shock has the order of $\xi$ $\leq$ (v$_{sh}$/c)$^2$ $\approx$ 10$^{-4}$ \citep{magic22}, so we consider $\xi_{-4}$ = 1 throughout this work. By using the model parameter values considered in this work (discussed in later sections) in equation \ref{eq13}, we find that the maximum energy of the electron population can go up to $\sim$ 30 GeV for the case in the study. This result is consistent with the idea adopted in this paper that the HE gamma-ray observed by \textit{Fermi}-LAT is produced from the IC cooling of the accelerated electrons.

On the other hand, similar to electrons, we also estimate the maximum energy the proton population can obtain. The timescale for the energy losses on pion production in inelastic \textit{p-p} collisions can be estimated as \citep{bednarek11},

\begin{equation}\label{eq13a}
    \tau_{pp} = (\sigma_{pp} k c n_{RG})^{-1} \approx 6.3 \times 10^4 \frac{r_{sh, 13}^2\:v_{sh, 8}}{\dot{M}_{RG, -4}}  s
\end{equation}

where $\sigma_{pp}$ $\approx$ 3 $\times$ 10$^{-26}$ cm$^2$ is the cross section for \textit{p-p} interaction, k = 0.5 is the inelasticity coefficient in this collision, r$_{sh, 13}$ = r$_{sh}$/(10$^{13}$ cm), v$_{sh, 8}$ = v$_{sh}$/(10$^8$ cm/s), and $\dot{M}_{RG, -4}$ = $\dot{M}_{RG}$/(10$^{-4}$ M$_{\odot}$ yr$^{-1}$). By equating the acceleration timescale of the protons $\tau^p_{acc}$ ($\approx$ 1E$_p$/$\xi_{-4}$B s) with the cooling timescale of the inelastic p-p interaction (equation \ref{eq13a}), we get the maximum proton energy \citep{bednarek11},

\begin{equation}\label{eq14}
    E^{max}_p \approx 63(\xi_{-4}B)\frac{r^2_{sh, 13} v_{sh, 8}}{\dot{M}_{RG, -4}}\: TeV
\end{equation}

Note that equation \ref{eq13a} is only valid if the condition $\dot{M}_{RG, -4}$ $>$ 0.2r$_{sh, 13}$v$^2_{sh, 8}$ is fulfilled. We find that the for the typical value of $\dot{M}_{RG}$ ($\approx$ 5 $\times$ 10$^{-7}$ M$_{\odot}$ yr$^{-1}$) considered in this paper, this condition is not fulfilled for all 4 days. In that case, the maximum energy of protons is determined by the escape along the shock, and can be estimated as \citep{bednarek11},

\begin{equation}\label{eq14a}
    E^{max}_p \approx 300(\xi_{-4}B)\frac{r_{sh, 13}}{v_{sh, 8}}\: TeV
\end{equation}

For the typical parameter values considered in this model, we find that the maximum energy obtained by the accelerated proton population can go up to $\sim$ 40 TeV during the first 4 days, following equation \ref{eq14a}. Due to the unavailability of the data in even higher energy regimes, in this work, we have used phenomenological cutoff energies in the form of E$^{cut}_e$ and E$^{cut}_p$ for electron and proton populations, respectively. We have only tuned the luminosities and cutoff energies of the electron and proton populations to explain the gamma-ray data in the HE-VHE range. Although the cutoff energies were varied to fit the data, we have considered that these cutoff energies can not be larger than the maximum energies of the electron and proton populations, as obtained by equation \ref{eq13} and \ref{eq14a}, respectively. One should note that since the TeV gamma-rays produced are within the radiation field of the companion RG star, the pair production process due to gamma-gamma absorption could be an additional mechanism in principle. However, the effect of the gamma-gamma absorption on the model spectrum is minor, as reported by \cite{hess22, magic22, diesing22}, so we do not include this process in our current theoretical modeling as its inclusion does not drastically alter the results presented in this work. We discuss this in detail in section \ref{sec4}. 

As discussed earlier, we also compare the synchrotron contribution of the model to the low radio frequency data observed by various radio observatories. But as studied in the case of the previous outburst of RS Oph in 2006 (studied during t $\sim$ 17 - 351 days post outburst at frequency $\nu$ = 0.24 - 1.4 GHz \citep{kantharia07}), the low frequency radio emission likely suffers from substantial foreground absorption due to the preexisting, ionized, warm, and clumpy red giant wind \citep{weiler02,kantharia07}. Since we use low frequency radio data in this case as well, we have also considered the effect of the absorption in this model. In fact, the radio emission at t $\sim$ 1 - 4 days occurs from the shocked shell further close to the WD+companion system, and will suffer more foreground absorption compared to relatively late time radio emission, as the radio emitting shell will be further away from the binary system at later times. The correction factors corresponding to the surrounding homogeneous, as well as clumpy mediums, were taken into account and were applied to the model synchrotron component using the following set of equations \citep{weiler02,kantharia07},

\begin{equation}\label{eq15}
\begin{aligned}
    F^{corr}_{sync} &= F_{sync} \left(\frac{t-t_0}{20\:days} \right)^{\beta} exp({-\tau^{CBM}_{homog}})\\ 
                    & \times \left[\frac{1 - exp({-\tau^{CBM}_{clumps}})}{\tau^{CBM}_{clumps}}\right]
\end{aligned}
\end{equation}

where,

\begin{equation}\label{eq16}
    \tau^{CBM}_{homog} = K_1 \left(\frac{\nu}{1\:GHz}\right)^{-2.1} \left(\frac{t-t_0}{20\:days} \right)^{\delta}
\end{equation}

and,

\begin{equation}\label{eq17}
    \tau^{CBM}_{clumps} = K_2 \left(\frac{\nu}{1\:GHz}\right)^{-2.1} \left(\frac{t-t_0}{20\:days} \right)^{\delta'}
\end{equation}

In this case, F$_{sync}$ and F$^{corr}_{sync}$ are model and absorption-corrected synchrotron fluxes, respectively, t - t$_0$ is the time since the outburst, K$_1$ is the attenuation by a homogeneous absorbing medium, and K$_2$ is the attenuation by a clumpy/filamentary medium at a frequency of 1 GHz, 20 days after the outburst. The optical depths $\tau^{CBM}_{homog}$ and $\tau^{CBM}_{clumps}$ are due to the ionized circumbinary material (CBM) external to the emitting region. The optical depths in the homogeneous and clumpy/filamentary CBM are described by $\delta$ and $\delta'$. Finally, $\beta$ signifies the rate of decline of flux density in the optically thin phase \citep{kantharia07}. The best-fit values of the parameters used in equations \ref{eq15}, \ref{eq16} and \ref{eq17} are, $\beta$ = -1.0311$^{+0.0322}_{-0.0310}$, K$_1$ = 0.0018$^{+0.0008}_{-0.0005}$, $\delta$ = -5.338$^{+0.2776}_{-0.2674}$, K$_2$ = 0.6349$^{+0.0941}_{-0.0956}$ and $\delta'$ = -2.9874$^{+0.0959}_{-0.1157}$, and they were obtained from fitting the low frequency radio light curve, constructed from the observational data corresponding to t $\sim$ 5 to 287 days post outburst at $\nu$ = 0.15 to 1.4 GHz (Nayana et al. in prep). The absorption-corrected model synchrotron flux, F$^{corr}_{sync}$, was used to compare the non-thermal synchrotron emission expected from our model to the observed radio data of the first 4 days. The results obtained from the model discussed in this section are given in the next section.

\section{Results} \label{sec3}

\begin{figure*}
    \centering
    \includegraphics[scale=0.5]{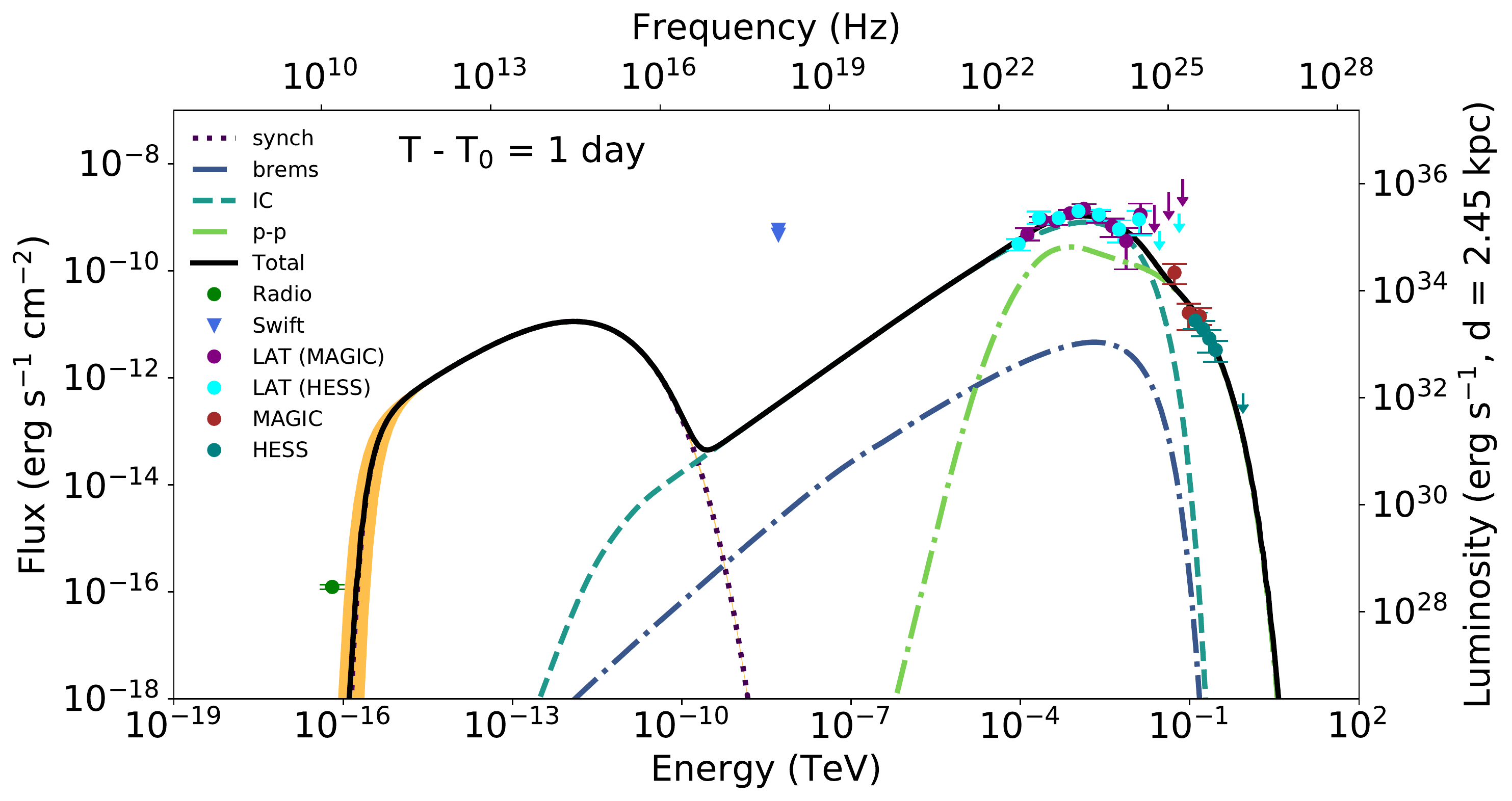}(a)
    \includegraphics[scale=0.5]{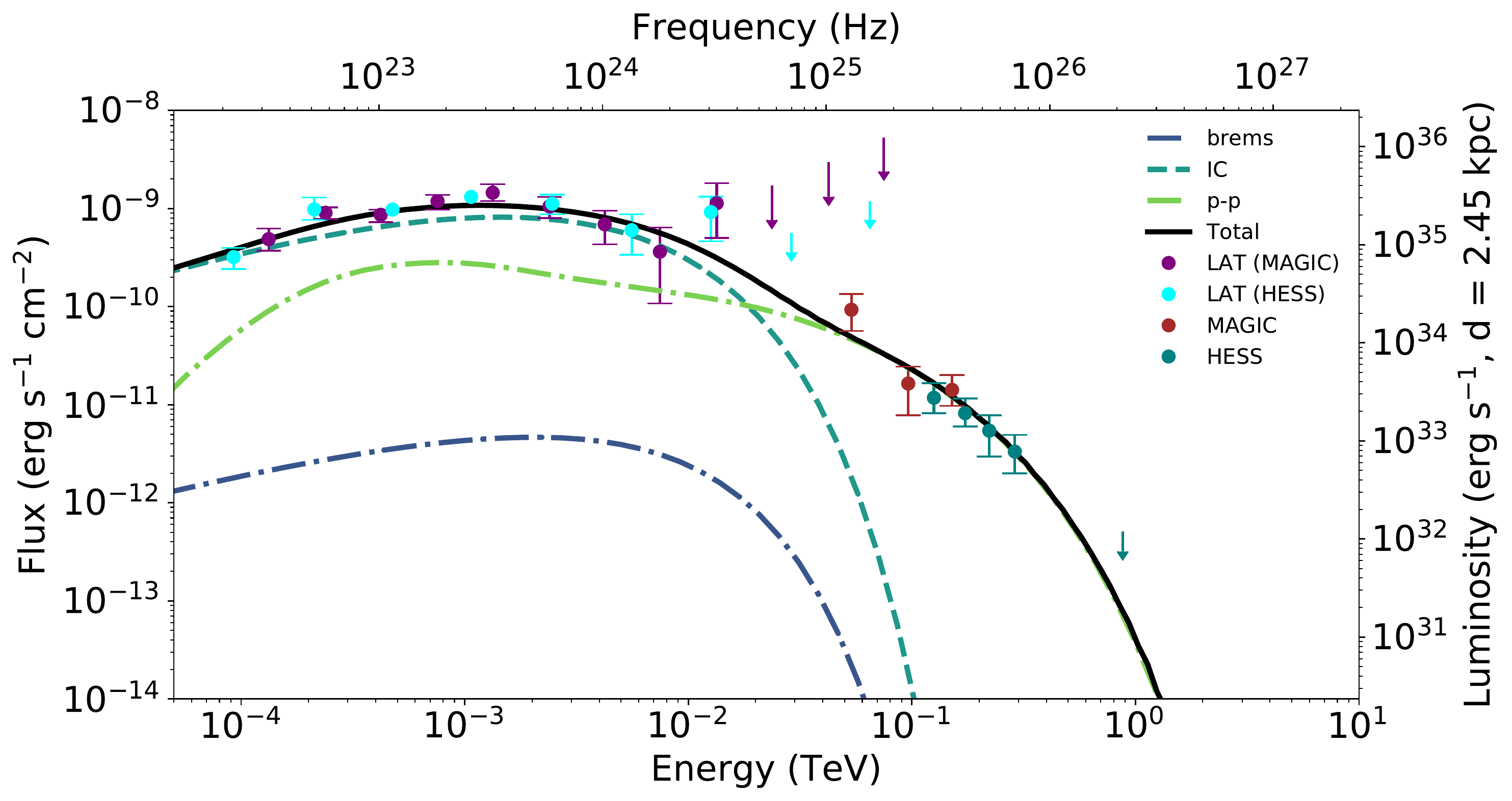}(b)
    \caption{The MWL SED data points for day 1 along with the SED computed from the model, are shown. The panel labeled (a) corresponds to the MWL SED corresponding to T - T$_0$ = 1 day after the outburst. The panel labeled (b) is the same as panel (a), but zoomed at HE-VHE range of the MWL SED. MAGIC (brown) and H.E.S.S. (teal) data points are taken from \cite{magic22}, and \cite{hess22}, respectively. \textit{Fermi}-LAT data points are taken from \cite{hess22} (cyan) and \cite{magic22} (purple). Radio data are shown in green. Swift-XRT model fluxes are shown in the form of upper limits \citep{page22} in royalblue triangles. The orange-shaded region corresponds to the uncertainty interval related to the parameters signifying absorption correction, as discussed in Section \ref{sec2}.}
    \label{fig1}
\end{figure*}

\begin{figure*}
    \centering
    \includegraphics[scale=0.5]{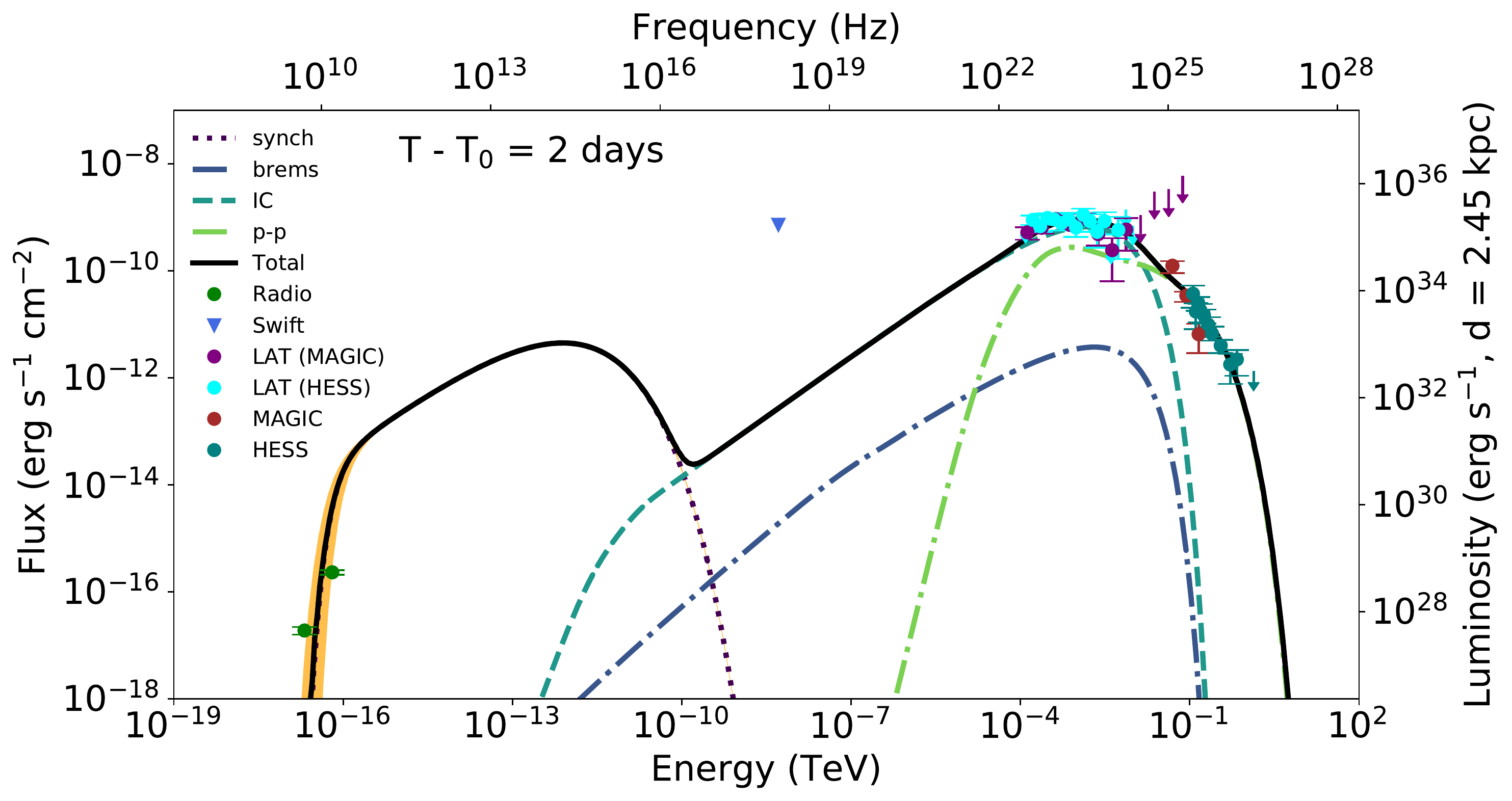}(a)
    \includegraphics[scale=0.5]{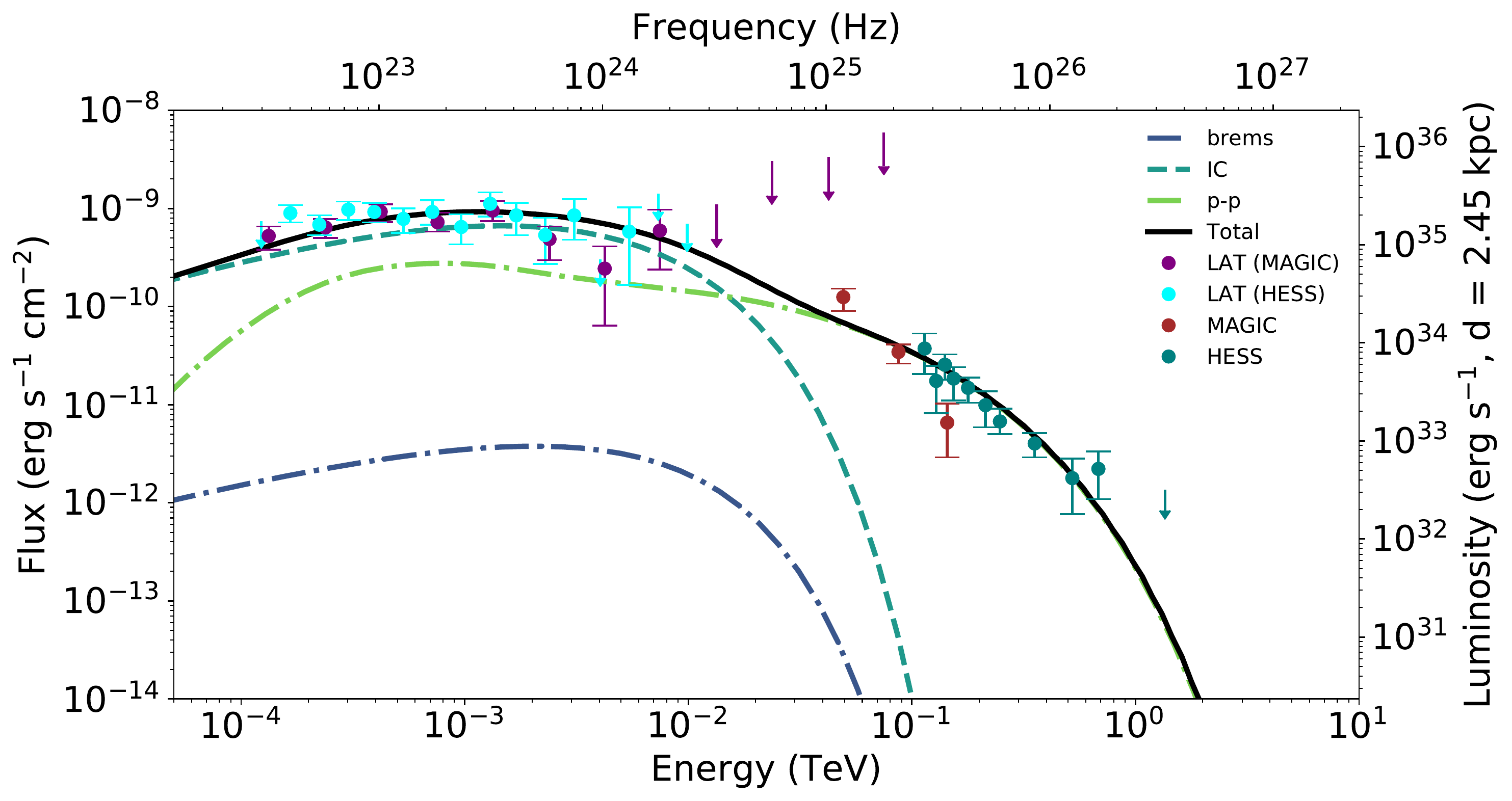}(b)
    \caption{The MWL SED data points for day 2 along with the SED computed from the model, are shown. The panel labeled (a) corresponds to the MWL SED corresponding to T - T$_0$ = 2 days after the outburst. The panel labeled (b) is the same as panel (a), but zoomed at HE-VHE range of the MWL SED. The color schemes of the plots are same as that given in Figure \ref{fig1}.}
    \label{fig2}
\end{figure*}

\begin{figure*}
    \centering
    \includegraphics[scale=0.5]{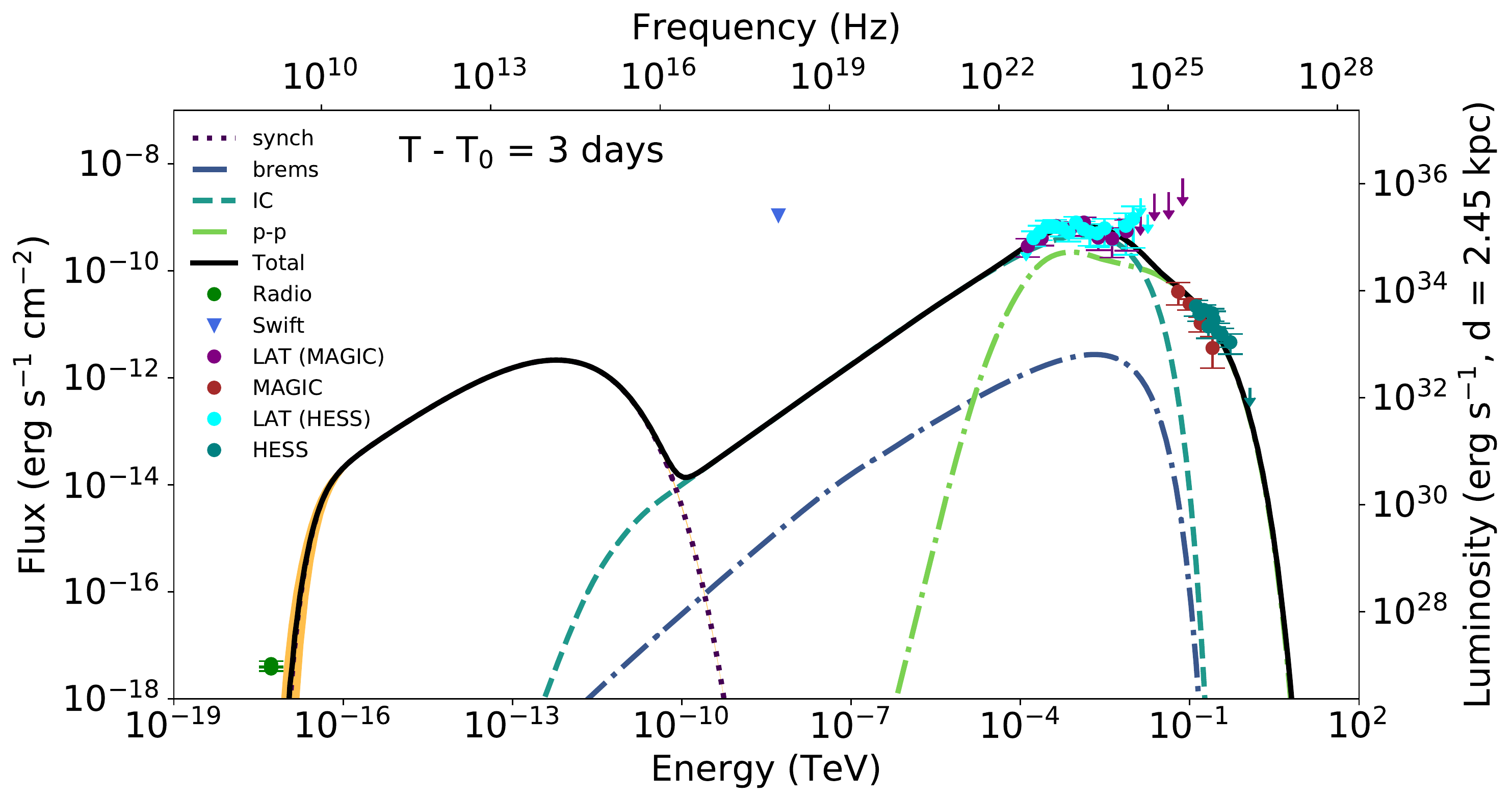}(a)
    \includegraphics[scale=0.5]{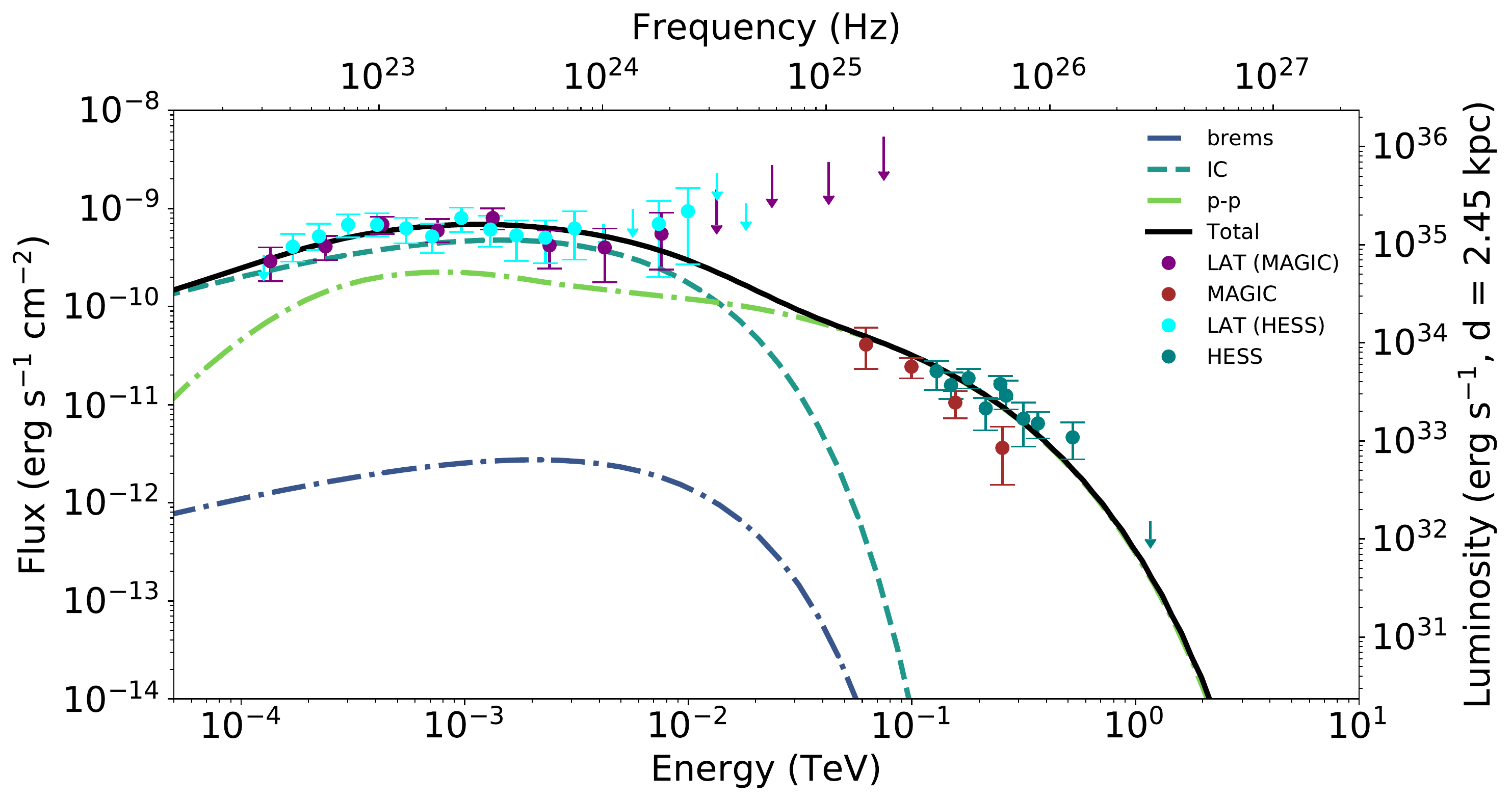}(b)
    \caption{The MWL SED data points for day 3 along with the SED computed from the model, are shown. The panel labeled (a) corresponds to the MWL SED corresponding to T - T$_0$ = 3 days after the outburst. The panel labeled (b) is the same as panel (a), but zoomed at HE-VHE range of the MWL SED. The color schemes of the plots are same as that given in Figure \ref{fig1}.}
    \label{fig3}
\end{figure*}

\begin{figure*}
    \centering
    \includegraphics[scale=0.5]{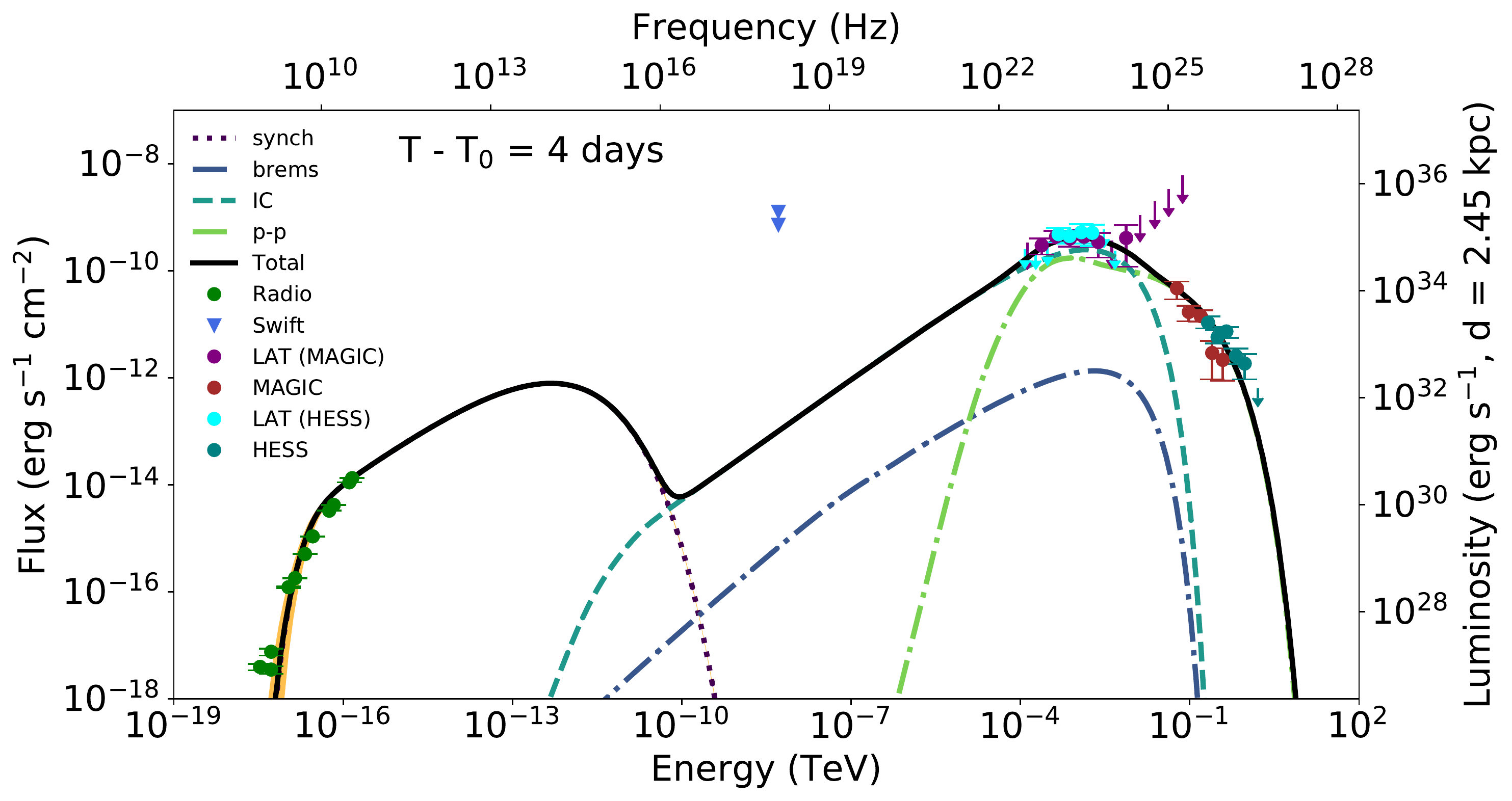}(a)
    \includegraphics[scale=0.5]{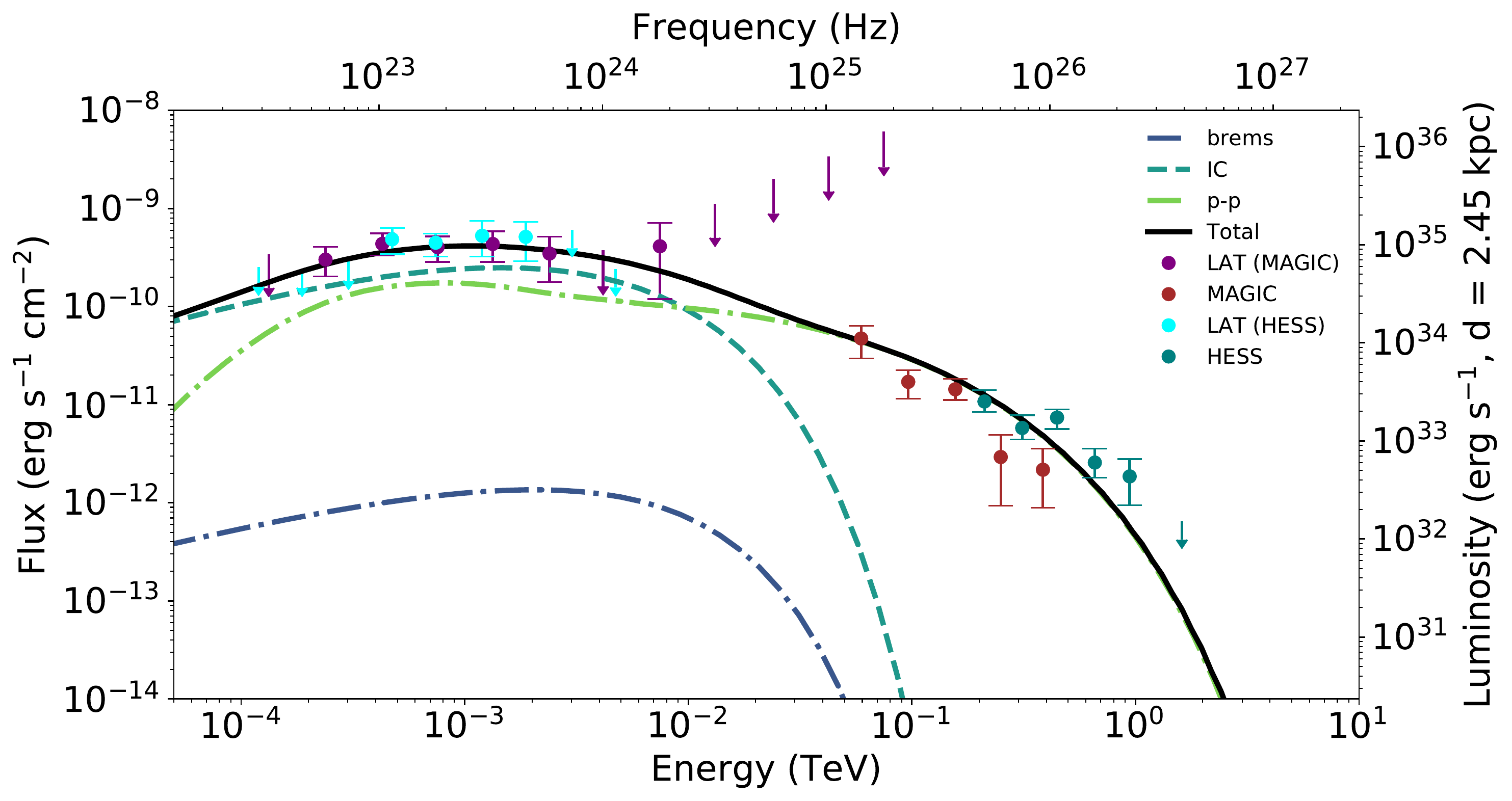}(b)
    \caption{The MWL SED data points for day 4 along with the SED computed from the model, are shown. The panel labeled (a) corresponds to the MWL SED corresponding to T - T$_0$ = 4 days after the outburst. The panel labeled (b) is the same as panel (a), but zoomed at HE-VHE range of the MWL SED. The color schemes of the plots are same as that given in Figure \ref{fig1}.}
    \label{fig4}
\end{figure*}

In this section, we discuss the results of our work. As discussed earlier, we only consider data from the first 4 days after the outburst. We have used the MAGIC, and H.E.S.S. data obtained by \cite{magic22} and \cite{hess22} to construct SEDs at the VHE range. The same authors have also analyzed the \textit{Fermi}-LAT data in their work. We have also used these \textit{Fermi}-LAT data points at the HE range. In the X-ray regime, \cite{page22} have analyzed data obtained from Swift X-ray Telescope (XRT), and provided the model X-ray flux data for the first 4 days, which we have plotted in our SEDs. However, these data points are model data points, and the model used in \cite{page22} is of thermal origin. So it is expected that our model, which gives non-thermal radiation as output, will not be able to explain these X-ray data points. For this reason, we have used the X-ray data points as upper limits in our SEDs to compare how the model SED compares with the X-ray data points. We have also used the first 4 days of radio data from different observatories such as Arcminute Microkelvin Imager - Large Array (AMI-LA), e - Multi-Element Radio Linked Interferometer Network  (e-MERLIN), MeerKAT and Karl G. Jansky Very Large Array (JVLA) to make multi-band SEDs. To our knowledge, this is the first time a nova eruption event has been studied in various wavebands (from radio to TeV energy range) simultaneously. The radio data points of the first 4 days, along with their frequencies, flux densities, and other information, are given in Table \ref{tab0}. The MWL SEDs, corresponding to 1 to 4 days after the outburst, are shown in Figures \ref{fig1}, \ref{fig2}, \ref{fig3}, and \ref{fig4}, respectively. The model parameters used to explain the MWL SEDs are given in Table \ref{tab1}.

\begin{table*}[]
    \centering
    \begin{tabular}{c c c c c}
    \hline
        Day & Frequency & Flux density & Observatory & References \\
            & (GHz) & (mJy) & & \\
        \hline
        Day 1 & 15.5 & 0.80 $\pm$ 0.08 & AMI-LA & \cite{ATel14849} (ATel 14849)\\
        \hline
        Day 2 & 5.0 & 0.38 $\pm$ 0.06 & e-MERLIN & \cite{ATel14849} (ATel 14849)\\
              & 15.5 & 1.50 $\pm$ 0.15 & AMI-LA & \cite{ATel14849} (ATel 14849)\\
        \hline
        Day 3 & 1.28 & 0.29 $\pm$ 0.03 & MeerKAT & \cite{ATel14849} (ATel 14849)\\
              & 1.28 & 0.349 $\pm$ 0.048 & MeerKAT & \cite{ruiter23}\\
        \hline
        Day 4 & 0.816 & 0.486 $\pm$ 0.068 & MeerKAT & \cite{ruiter23}\\
         & 1.28 & 0.276 $\pm$ 0.045 & MeerKAT & \cite{ruiter23}\\
         & 1.28 & 0.593 $\pm$ 0.087 & MeerKAT & \cite{ruiter23}\\
         & 2.6 & 4.710 $\pm$ 0.140 & JVLA & \cite{ATel14886} (ATel 14886)\\
         & 3.4 & 5.328 $\pm$ 0.093 & JVLA & \cite{ATel14886} (ATel 14886)\\
         & 5.1 & 10.012 $\pm$ 0.058 & JVLA & \cite{ATel14886} (ATel 14886)\\
         & 7.0 & 15.607 $\pm$ 0.057 & JVLA & \cite{ATel14886} (ATel 14886)\\
         & 13.7 & 24.088 $\pm$ 0.055 & JVLA & \cite{ATel14886} (ATel 14886)\\
         & 16.5 & 25.573 $\pm$ 0.059 & JVLA & \cite{ATel14886} (ATel 14886)\\
         & 31.1 & 35.973 $\pm$ 0.140 & JVLA & \cite{ATel14886} (ATel 14886)\\
         & 34.9 & 38.318 $\pm$ 0.160 & JVLA & \cite{ATel14886} (ATel 14886)\\
        \hline 
    \end{tabular}
    \caption{Low frequency radio data used in this work.}
    \label{tab0}
\end{table*}

Figures \ref{fig1}, \ref{fig2}, \ref{fig3}, and \ref{fig4} show that the VHE MAGIC and H.E.S.S. gamma-ray data points can be explained by gamma-rays produced by the hadronic interaction between protons accelerated in the forward shock, and protons present in the RG wind. This conclusion is similar to that explored in \cite{magic22, hess22, zheng22}. However, we find that the hadronic component does not entirely explain the HE gamma-ray data points for the choices of parameter values considered in our model. Nevertheless, our model is similar to that discussed in \cite{sitarek12, magic15}, and moreover, the observed VHE gamma-ray data better constrain our model. The proton luminosity and the cutoff energy were varied to explain the VHE MAGIC and H.E.S.S. data. Similar to the protons, we vary the electron luminosity and cutoff energy to explain the HE gamma-ray data. We also considered gamma-rays originating from Bremsstrahlung emission, where accelerated electrons interact with the RG wind, but we found it not to be significant enough to contribute to the HE-VHE gamma-ray regime. From Figures \ref{fig1}, \ref{fig2}, \ref{fig3}, and \ref{fig4}, we can see that the combination of leptonic and hadronic components from the nova explains the HE-VHE gamma-ray data well for all 4 days of observations. From Table \ref{tab1}, one can also see that the electron-to-proton (L$_e$/L$_p$) luminosity ratio obtained from our model contradicts the condition presented in \cite{magic15}, in which a limit of L$_p$ $\leq$ 0.15 L$_e$ was given, considering MAGIC upper limits available at that time. But using MAGIC and H.E.S.S. data, we were able to show that proton luminosity is, in fact, a few orders higher than electron luminosity in the case of a lepto-hadronic model, contrary to what was obtained in \cite{magic15}. 

Although not all recurrent novae are detectable in synchrotron radio, previous RS Oph outburst in 2006 showed clear non-thermal origin until 351 days post outburst at $\nu$ = 0.24 to 1.4 GHz \citep{kantharia07, kantharia16}. The low frequency radio light curve of RS Oph 2006 outburst has been explained using a phenomenological model of non-thermal synchrotron emission, while taking into account the absorption due to the clumpy red giant wind \citep{kantharia07}. The clumpiness of the ambient medium leads to a decrement of the optical depth encountered by the synchrotron photon compared to that of a uniform medium, and hence an increment in the observed synchrotron flux can be expected \citep{kantharia16}. The radio light curve of the 2021 RS Oph outburst also hints towards a non-thermal origin, at least until 287 days post outburst at $\nu$ = 0.15 to 1.4 GHz, as will be discussed in Nayana et al. (in prep). Consequently, we have considered the non-thermal, synchrotron emission expected from our model and compared it with the observed radio data in low frequencies. After calculating the total synchrotron component using the same parent electron spectrum considered to explain the HE \textit{Fermi}-LAT gamma-ray data, we have also corrected it by considering absorption of the synchrotron photons due to homogenous, as well as clumpy medium present in the line of sight, as the clumpiness of the medium affects the synchrotron emission in the case of the 2021 RS Oph outburst as well. As seen from Figures \ref{fig1}, \ref{fig2}, \ref{fig3}, and \ref{fig4}, we find that the absorption-corrected model synchrotron emission is consistent with all 4 days of radio data.  Through the MWL modeling reported in this paper, we could show that a non-thermal synchrotron component is responsible for the low frequency radio emission at the early stages of RS Oph nova outburst. Since a partially thermal contribution could be a possibility, we have further tested the non-thermal origin of the low radio frequency data by calculating the brightness temperature using the following equation \citep{chomiuk21},

\begin{equation}\label{eq18}
    \frac{T_B}{K} = 1200 \left( \frac{S_\nu}{mJy} \right) \left( \frac{\nu}{GHz} \right)^{-2} \left( \frac{\theta}{arcsec} \right)^{-2}
\end{equation}
where, S$_{\nu}$ is the observed flux density at frequency $\nu$. We have considered S$_{\nu}$ and $\nu$ from Table \ref{tab0}, and $\theta$ was calculated using the distance, shock velocity, and expansion time (i.e., the shock radius) from Table \ref{tab1}, for all 4 days after the outburst. We have found a very high brightness temperature for all 4 days, i.e., T$_B$ $\geq$ 10$^5$ - 10$^{6}$ K, which is indicative of shock-induced, non-thermal synchrotron emission \citep{chomiuk21}. The presence of synchrotron emission is expected from shock interaction, for novae like RS Oph, V745 Sco, V3890 Sgr, since these source systems have a RG star as their companion. On the other hand, novae like U Scorpii has an evolved K2 IV companion, while the companion in T Pyxidis is a main sequence star. Consequently, synchrotron radio emission from these sources were not detected in low frequency radio surveys \citep{anupama13, pavana19}.

\section{Discussion} \label{sec4}

In this paper, we have provided a single shock, multiple particle population, lepto-hadronic model of RS Oph nova outburst and used this model to explain the HE-VHE gamma-ray data observed for the first 4 days. In addition, we have also considered simultaneous low-frequency radio observations (given in Table \ref{tab0}) and used that to test the presence of accelerated electrons in the source region. Time-dependent conditions were considered for the components of the ambient medium, following \cite{magic22} and Nayana et al. (in prep). By solving the TDDL equation (equation \ref{eq7}), we explain the MWL data ranging from highest energy gamma-rays to low energy radio band. We summarize the key points of this work below.

\begin{enumerate}

\item Considering only the data in the HE-VHE range, it is difficult to confirm whether a purely hadronic or leptonic scenario is at play in the source region. Although both IC emission in a leptonic scenario and inelastic \textit{p-p} interaction in a hadronic scenario is able to explain the HE emission, when VHE emission is also considered, the hadronic scenario gets an advantage, since leptonic interaction is affected by Klein-Nishina suppression at VHE range. However, it is likely that electrons also get accelerated, similar to protons. So a reasonable contribution from electrons to the total gamma-ray flux should be expected. This fact suggests that HE gamma-ray data is more likely to be explained by the leptonic component, whereas VHE gamma-ray data is hadronic in origin. The lepto-hadronic interpretation of RS Oph nova outburst discussed in this paper, not only satisfactorily explains first 4 days of MWL data, such a scenario can be used to explain the temporal features of the outburst. For example, when fitted with a power law, the index for the temporal decay in the case of H.E.S.S. data came out to be $\alpha_{HESS}$ = 1.43 $\pm$ 0.18, whereas the same for \textit{Fermi}-LAT is $\alpha_{LAT}$ = 1.31 $\pm$ 0.07 \citep{hess22}. Although both of these values are roughly consistent within error, different best-fit values may indicate the different origins of HE and VHE gamma-ray data, which can potentially be solved by a lepto-hadronic scenario. In addition to the points discussed above, the gamma-ray light curve peak observed by H.E.S.S. in TeV energies is delayed by a few days as compared to the GeV gamma-ray light curve peak observed by \textit{Fermi}-LAT, as the \textit{Fermi}-LAT light curve peaks on 2021 Aug 9–10 \citep{fermi22}, and the H.E.S.S. light curve peaks on 2021 Aug 12 \citep{hess22}. \cite{hess22} have explained the observed time delay between light curve peaks at different energies as the finite acceleration time of the $>$ 1 TeV protons in a single shock, single particle population scenario. On the other hand, \cite{diesing22} have explained the same by considering a slow, highly luminous shock component, which produces the GeV emission at early times, and a fast, less luminous shock component, which produces hardened TeV emission at later times. In the alternate model discussed in this paper, although we take a single shock scenario, the GeV and TeV emissions have been attributed to electron and proton populations, respectively. The IC cooling timescale of electrons is less than that of \textit{p-p} interaction, and the maximum energy of the electron population accelerated at the shock can go up to $\sim$ 30 GeV, as discussed in section \ref{sec2}. Consequently, electrons will likely lose energy very efficiently through IC cooling, and the bulk of the gamma-rays produced will be observed in the GeV range at early times. On the other hand, the hadronic interaction timescale is comparatively higher than the IC cooling timescale; hence it will take more time for the accelerated proton population to produce observable gamma-rays. In addition, the maximum energy of the proton population can go up to $\sim$ 40 TeV (see section \ref{sec2}); hence most of the gamma-rays produced through inelastic \textit{p-p} interaction will be observed in the TeV range at slightly later times. So in the single shock, lepto-hadronic scenario discussed in this paper, the TeV delay can be naturally explained by attributing the GeV light curve peak to the fast cooling of the electron population and the TeV light curve peak to the hadronic interaction of the proton population. 
Nevertheless, in this work, we refrain from discussing the temporal features of the RS Oph outburst in more detail, and primarily focus on explaining the spectral features of the early time MWL data with a lepto-hadronic model, similar to that explored in \cite{sitarek12}.

\item From the discussion above, it is evident that a single shock, lepto-hadronic scenario, can explain the HE-VHE gamma-ray data obtained from RS Oph nova outburst, comparatively better than a single shock, purely hadronic or a purely leptonic model. Nevertheless, testing the same lepto-hadronic scenario in an MWL context should provide a more compelling argument for the case. To that end, as discussed before, we have also included the results from simultaneous low frequency radio observations (Table \ref{tab0}), in conjunction with HE-VHE gamma-ray data, to construct MWL SEDs of RS Oph. From our model, we were able to show that the synchrotron component obtained from the parent electron population is consistent with the observed radio data for all 4 days. The magnetic field near the forward shock used to calculate the synchrotron emission was fixed a priori, so the fact that synchrotron emission from the parent electron population was able to explain all 4 days of the radio data indicates that the presence of accelerated parent electron population near the shock region is a valid assumption. We note that in a recent paper by \cite{ruiter23}, the authors posited that an old synchrotron emission component from 2006 RS Oph outburst may be responsible for the early-time (t $<$ 5 days) low frequency radio emission observed. However, in our model, we show that the synchrotron emission component resulting from electrons, freshly accelerated during the 2021 RS Oph outburst, can also consistently explain the early-time, low frequency radio emission. 

\item Modeling VHE gamma-ray emission using a hadronic scenario inevitably leads to the production of astrophysical neutrinos. Detecting these astrophysical neutrinos is crucial since that would confirm the presence of hadronic interaction at the VHE range. High energy neutrinos from symbiotic nova such as V407 Cygni has been predicted earlier \citep{soeb10}. To calculate the flux of the muonic neutrinos $\nu_{\mu}$ + $\Tilde{\nu}_{\mu}$ resulting from our model, we use the semi-analytical formulation developed in \cite{kelner}. Following \cite{kelner}, we have included the muonic neutrinos produced from direct decay of charged pions ($\pi$ $\rightarrow$ $\mu$ $\nu_{\mu}$), and from the decay of muons ($\mu \rightarrow$ e $\nu_{\mu}$ $\nu_e$). Our estimated muon neutrino flux is shown in Figure \ref{fig5}.

IceCube, ANTARES, and KM3NeT/ARCA are state-of-the-art neutrino detectors capable of detecting astrophysical neutrinos from Galactic sources. IceCube has searched for muon neutrino flux from the direction of RS Oph during a three day time window covering the beginning of observed optical outburst. A time-integrated muon neutrino flux upper limit of 4.8 $\times$ 10$^{-5}$ TeV cm$^{-2}$ at 90$\%$ confidence level was derived, under the assumption of E$^{-2}$ power law, between 2 TeV and 10 PeV \citep{pizzuto21}. Our estimated muon neutrino flux shown in Figure \ref{fig5} is well below the derived IceCube upper limit. KM3NeT/ARCA is a next-generation neutrino observatory that can detect neutrinos across a wide declination range \citep{aiello19}. Considering RS Oph declination of -06$^{o}$42$'$28$''$, we find that although the source position is well within the observable range of KM3NeT/ARCA, the maximum neutrino flux obtained from RS Oph in our model is well below the sensitivity of KM3NeT/ARCA at that declination. Note that MGRO J1908+06 is the closest to RS Oph in terms of declination, amongst the sources reported in \cite{aiello19}. KM3NeT/ARCA predicted neutrino detection from MGRO J1908+06 after 6 years of observation. However, MGRO J1908+06 is a continuously emitting source, and from Figure 2 of \cite{aiello19}, it can be seen that the neutrino flux of the MGRO source is $\approx$ 10$^{-12}$ erg cm$^{-2}$ s$^{-1}$ at 10 TeV. Since nova such as RS Oph is a transient source, i.e., it can only be observed during a few months timescales, and the neutrino flux at 10 TeV is negligible for RS Oph, as can be seen from Figure \ref{fig5}, it is very difficult to observe astrophysical neutrinos from RS Oph with even the next generation neutrino detectors. \cite{magic22,guetta22} also corroborated that neutrino emission is not expected to be detected from the novae outbursts such as RS Oph by the current experiments. With neutrino being a by-product of the hadronic interaction, the neutrino detection would have been a smoking gun evidence for the hadronic scenario in the source region. Neutrino flux measurements from a future higher magnitude nova outburst will be able to confirm the pure or partial hadronic nature of particle acceleration.

\end{enumerate}

We note here that in this work, we do not intend to ``fit'' the data, as the MWL data, especially in radio and X-ray ranges, is poorly constrained. Consequently, we have only adopted different parameter values following previous literature, reported in Table \ref{tab1}, to explain the MWL SEDs corresponding to 1 to 4 days after the ourburst, and further, to discuss the conditions necessary for a lepto-hadronic interpretation of 2021 RS Oph outburst. Nevertheless, we have tried to ``fit'' the data by treating 7 parameters of the model ( i.e., L$_p$, $\alpha_p$, E$^{cut}_p$, L$_e$, $\alpha_e$, E$^{cut}_e$, B)  as free parameters instead of adopting the values reported in Table \ref{tab1}, for 4 MWL SEDs given in Figures \ref{fig1}, \ref{fig2}, \ref{fig3}, and \ref{fig4}. As expected, we have found that, although the best-fit values obtained from such method are very close to the adopted values reported in Table \ref{tab1}, the 1$\sigma$ errors of the free parameters are very large, due to the small number of constraining data points at the early time. As a result, we refrain from reporting the fit results, and their 1$\sigma$ uncertainties in the paper. Future simultaneous observations in multiple wavelengths will be helpful in constraining the model parameters, thereby further confirming the emission mechanism occuring in novae outbursts.

A hard spectral index ($\sim$ 1.5) for the parent electron spectrum was previously used in the lepto-hadronic interpretations of V407 Cygni and V339 Del novae outbursts \citep{sitarek12, martin13, magic15}. Since this work involves a phenomenological modeling of RS Oph 2021 outburst, we have assumed a similar hard electron index as that considered in previous literature. We were able to show that emission from parent electron population with hard spectral index, is able to explain the radio and HE gamma-ray data, which essentially proves to be a natural continuation of previous studies done in explaining novae outbursts with a lepto-hadronic scenario. The plausible acceleration scenario behind this hard index is beyond the scope of this paper. Nevertheless, in general, magnetic reconnections in Poynting flux dominated outflow can produce $\sim $ 1.5 spectral index. Another possibility is the electrons originating from proton-photon interactions, which can have a very hard spectral index.

It has been found in multiple studies previously done, that the RG companion star, being a  M-type star, has an effective surface temperature in the range of 3200 K $<$ T $<$ 4400 K \citep{pavlenko08, pavlenko16, barry07}. Given the temperature at the surface of the RG companion, the temperature of the dense wind produced from the same companion is expected to be less than, or at most of the same order of, the surface temperature. As discussed earlier, we have additionally assumed that the relevant conditions related to the wind is considered before the wind is heated and ionized by the outburst, similar to that considered in \cite{chomiuk12} for nova V407 Cygni, in which a wind temperature of $\sim$ 700 K was set. Taking these factors into account, we have considered an order of magnitude estimate by assuming the RG wind temperature of 10$^3$ K in the equipartition magnetic field, given by equation \ref{eq6}. The order of the resulting magnetic field is not only comparable with that found in previous literature \citep{martin13}, but also, the resulting synchrotron emission due to this magnetic field explains the observed radio emission consistently. A somewhat higher value of the RG wind temperature of 10$^4$ K was adopted in \cite{bode85}, but using this value of the wind temperature would result in a larger magnetic field that is neither consistent with the observations, nor with previous estimates.

The magnetic field near the shock region can also be amplified by the streaming of accelerated particles in the shock precursor. The amplified magnetic field can be proportional to the equipartition magnetic field, given by equation \ref{eq6}. This would introduce a proportionality factor f$_B$, which would essentially act as a free parameter, considering the equipartition magnetic field to be fixed. However, as discussed earlier, it is very difficult to constrain the total magnetic field with such less number of degrees of freedom in low energies (no data in X-ray, few data in radio). Consequently, it will be difficult to constrain the value of f$_B$ , in turn, due to poor statistic in low energies. Considering a value of f$_B$ associated with magnetic field without properly constraining it, would turn out be an ad-hoc assumption, as no typical values of f$_B$ are known a priori. Since our main motivation behind this work is to report the conditions necessary to explain the MWL emission with a lepto-hadronic scenario, we have tried to fix multiple parameter values beforehand, following some reasonable assumptions. As a result, in this case, we refrain from introducing f$_B$ as another free parameter of the model, and fix the magnetic field to the
values estimated by the equipartition argument to avoid further complications. 

\cite{magic22} has noted that it is difficult to explain the shape of curvature observed in the measured gamma-ray spectrum between 50 MeV to 250 GeV range with leptonic processes. It is indeed the case that a single IC leptonic component will be unable to explain the observed curvature, as well as the VHE gamma-ray data, without introducing a strong break in the parent electron spectrum. However, in the lepto-hadronic model discussed in this paper,  the curvature is explained by a combined contribution of the leptonic and the hadronic components (mainly dominated by the leptonic component). We have used an electron injection spectrum with cutoff at 10s of GeV, which is also limited by the maximum attainable energy of the electrons. So no ad-hoc break was needed in the electron spectrum. By combining the resulting leptonic component with the hadronic component used to explain the VHE gamma-ray data, the curvature has been automatically explained. We have provided zoomed-in images of only the HE-VHE gamma-ray section of the MWL SED plots, to show that the total combined model SEDs (primarily dominated by the leptonic component) readily explains the curvature observed in 50 MeV to 250 GeV range (please see Figures \ref{fig1}, \ref{fig2}, \ref{fig3}, and \ref{fig4}).

We have neglected the effect of gamma-gamma attenuation in this model, following the conclusion drawn by previous studies on RS Oph \citep{hess22, magic22, diesing22}. \cite{hess22} have posited that the attenuation of TeV gamma-rays occurs due to the interaction with optical–IR photons, and X-rays provide the dominant target for GeV photons. Gamma-gamma attenuation is most relevant for VHE gamma-ray, whereas, it is important for GeV photons only after few hours after the outburst, when the source size is small. This is because in early time, optical-IR luminosity is larger ($\sim$ 10$^{37}$ erg/s) \citep{kafka}, compared that for X-ray ($\sim$ 10$^{35}$ erg/s) at a distance of 2.45 kpc \citep{sokolski06}. \cite{hess22} have shown that for the optical depth for photons considered in their work (see equation S31 of the supplementary material, \cite{hess22}), the spectrum change due to the gamma-gamma absorption is expected to be minor (please see Figures S10A and S10B of the supplementary material, \cite{hess22}). \cite{magic22} have also stated that the effect of gamma-gamma attenuation of the emission in the photosphere radiation field of RS Oph is not significant. \cite{diesing22} has posited that although close to T - T$_0$ $\simeq$ 1 day, the gamma-gamma attenuation is expected to be modest (by a factor of $\sim$ 2),  this attenuation is negligible at the radius corresponding to day 1 to 4, i.e., during the rise and subsequent peak of the observed TeV luminosity. Consequently, the authors also neglected absorption in their emission estimates. Following the arguments presented above, it can be assumed that the effect of gamma-gamma attenuation is not significant from the 1st day onward after the outburst. Nevertheless, even a modest gamma-gamma absorption is also unlikely to change the main conclusion of this paper. Consequently, we have chosen to neglect the effect of gamma-gamma absorption in this particular work. 

It is to be noted that the source region was considered to be spherically symmetric for the modeling of the first 4 days after the outburst in this work. But given the binary separation of 1.48 AU ($\approx$ 2.2 $\times$ 10$^{13}$ cm) and the shock radii given in Table \ref{tab1}, one can see that the binary separation and shock radius values are comparable, at least for first 4 days. In such conditions, the density profile tends to become more complicated, given that an anisotropic distribution better represents the ambient matter at this stage. In the first few days, the shock is more likely to expand as a bipolar blast wave moving orthogonal to the accretion disk \citep{hess22}. Modeling this bipolar (later quasi-spherical) shock wave, centering at the WD position, is very complicated, sophisticated, and beyond the scope of this paper. However, as followed in \cite{zheng22}, we expect the spherical shock assumption to be a good approximation, and the inclusion of an anisotropical shock treatment will not significantly change the main aim of the results reported in this paper.

\section{Conclusion} \label{sec5}

In conclusion, we have provided a theoretical lepto-hadronic interpretation of MWL data observed from the 2021 RS Oph nova outburst. We have shown that the VHE gamma-ray data can be explained by a hadronic component, whereas the HE gamma-ray data is satisfied with a leptonic component. The presence of a leptonic component also helps explain the low frequency radio data points observed for all 4 days. Our work vastly improves upon the inferences drawn by \cite{sitarek12, magic15}, and better interprets the MWL data compared to a purely hadronic or a purely leptonic scenario discussed in \cite{magic22, hess22}. However, further observations are needed to categorically confirm the proper nature of particle acceleration occurring in nova's environment. Given that recurrent outbursts of novae are relatively regular occurrence in the Galaxy, future observations of HE-VHE gamma-rays with current generation observatories (\textit{Fermi}-LAT, MAGIC, H.E.S.S.), as well as next-generation gamma-ray observatories such as Cherenkov Telescope Array \citep[CTA;][]{CTA19} and Southern Wide-field Gamma-ray Observatory \citep[SWGO;][]{SWGO19}, along with further simultaneous observations in radio and X-ray energy ranges, will be able to confirm the exact nature of the emission. Moreover, sophisticated simulations taking into account the anisotropic matter distribution and an aspherical shock wave structure would also be helpful for complete theoretical modeling of the nova region. If next-generation neutrino observatories can detect neutrinos from the outburst site, then effective contributions of hadronic and leptonic interactions will also be unveiled, thus opening new possibilities to model and understand the exact nature of this type of interesting source.  

\begin{figure}[ht]
    \centering
    \includegraphics[width=0.5\textwidth]{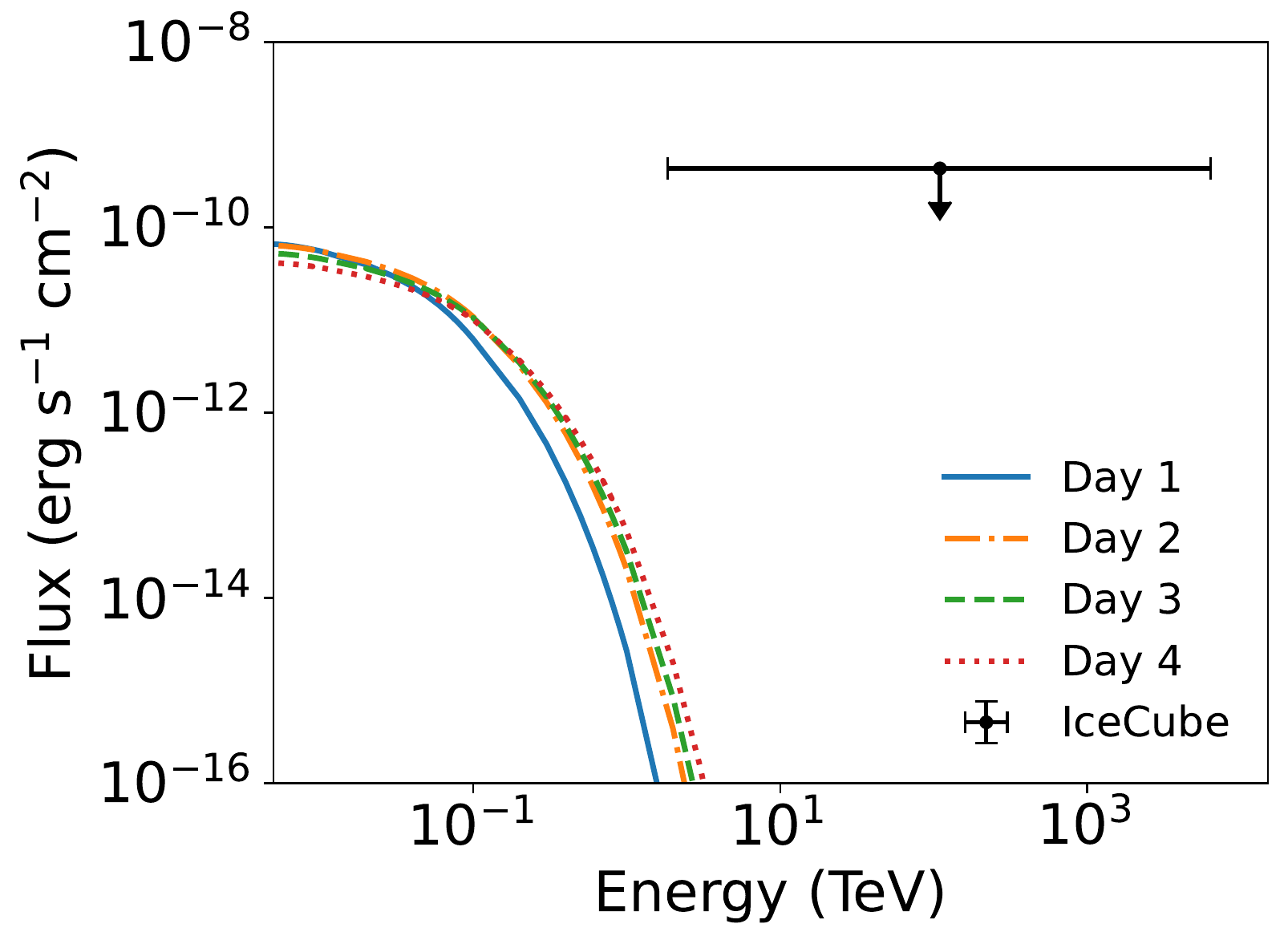}
    \caption{The estimated total muonic neutrino flux reaching the Earth from RS Oph, according to the lepto-hadronic model explored in this work. 90$\%$ C.L. upper limit obtained by IceCube is given in black \citep{pizzuto21}.}
    \label{fig5}
\end{figure}


\begin{deluxetable*}{ccccccccc}
\tablecaption{\label{tab1} Parameters used in the model.}
\tablewidth{0pt}
\tablehead{
\colhead{Day} & \colhead{Component} & \colhead{Parameter} & \colhead{Value}}
\startdata
Day 1 & Nova structure & Shock velocity (v$_{sh}$) & 4.5 $\times$ 10$^8$ cm/s\\
               & & Shock radius (r$_{sh}$) & 4 $\times$ 10$^{13}$ cm\\
      & Hadronic & Injection spectral index ($\alpha_p$) & 2.2\\
               & & Cutoff energy (E$^{cut}_p$) & 400 GeV\\
               & & Luminosity (L$_p$) & 2.14 $\times$ 10$^{35}$ erg/s\\
               & & Ejecta density (n$_{ej}$) & 1.62 $\times$ 10$^{10}$ cm$^{-3}$\\
               & & RG wind density (n$_{RG}$) & 9.9 $\times$ 10$^8$ cm$^{-3}$\\
      & Leptonic & Injection spectral index ($\alpha_e$) & 1.5\\
                 & & Cutoff energy (E$^{cut}_e$) & 10 GeV\\
                 & & Luminosity (L$_e$) & 3.5 $\times$ 10$^{32}$ erg/s\\
                 & & Magnetic field (B) & 0.11 G\\
                 & & Photon energy density (u$_{ph}$) & 1.26 erg cm$^{-3}$\\
                 & & Electron-to-proton luminosity ratio (L$_e$/L$_p$) & 1.6 $\times$ 10$^{-3}$\\
\hline
Day 2 & Nova structure & Shock velocity (v$_{sh}$) & 3.5 $\times$ 10$^8$ cm/s\\
               & & Shock radius (r$_{sh}$) & 6.34 $\times$ 10$^{13}$ cm\\
      & Hadronic & Injection spectral index ($\alpha_p$) & 2.2\\
               & & Cutoff energy (E$^{cut}_p$) & 600 GeV\\
               & & Luminosity (L$_p$) & 5.5 $\times$ 10$^{35}$ erg/s\\
               & & Ejecta density (n$_{ej}$) & 4.05 $\times$ 10$^{9}$ cm$^{-3}$\\
               & & RG wind density (n$_{RG}$) & 3.9 $\times$ 10$^8$ cm$^{-3}$\\
      & Leptonic & Injection spectral index ($\alpha_e$) & 1.5\\
                 & & Cutoff energy (E$^{cut}_e$) & 10 GeV\\
                 & & Luminosity (L$_e$) & 7.2 $\times$ 10$^{32}$ erg/s\\
                 & & Magnetic field (B) & 0.07 G\\
                 & & Photon energy density (u$_{ph}$) & 0.5 erg cm$^{-3}$\\
                 & & Electron-to-proton luminosity ratio (L$_e$/L$_p$) & 1.3 $\times$ 10$^{-3}$\\
\hline
Day 3 & Nova structure & Shock velocity (v$_{sh}$) & 3.1 $\times$ 10$^8$ cm/s\\
               & & Shock radius (r$_{sh}$) & 8.32 $\times$ 10$^{13}$ cm\\
      & Hadronic & Injection spectral index ($\alpha_p$) & 2.2\\
               & & Cutoff energy (E$^{cut}_p$) & 700 GeV\\
               & & Luminosity (L$_p$) & 7.6 $\times$ 10$^{35}$ erg/s\\
               & & Ejecta density (n$_{ej}$) & 1.8 $\times$ 10$^{9}$ cm$^{-3}$\\
               & & RG wind density (n$_{RG}$) & 2.3 $\times$ 10$^8$ cm$^{-3}$\\
      & Leptonic & Injection spectral index ($\alpha_e$) & 1.5\\
                 & & Cutoff energy (E$^{cut}_e$) & 10 GeV\\
                 & & Luminosity (L$_e$) & 8.8 $\times$ 10$^{32}$ erg/s\\
                 & & Magnetic field (B) & 0.05 G\\
                 & & Photon energy density (u$_{ph}$) & 0.29 erg cm$^{-3}$\\
                 & & Electron-to-proton luminosity ratio (L$_e$/L$_p$) & 1.1 $\times$ 10$^{-3}$\\
\hline
Day 4 & Nova structure & Shock velocity (v$_{sh}$) & 2.8 $\times$ 10$^8$ cm/s\\
               & & Shock radius (r$_{sh}$) & 1 $\times$ 10$^{14}$ cm\\
      & Hadronic & Injection spectral index ($\alpha_p$) & 2.2\\
               & & Cutoff energy (E$^{cut}_p$) & 850 GeV\\
               & & Luminosity (L$_p$) & 9.2 $\times$ 10$^{35}$ erg/s\\
               & & Ejecta density (n$_{ej}$) & 1.05 $\times$ 10$^{9}$ cm$^{-3}$\\
               & & RG wind density (n$_{RG}$) & 1.5 $\times$ 10$^8$ cm$^{-3}$\\
      & Leptonic & Injection spectral index ($\alpha_e$) & 1.5\\
                 & & Cutoff energy (E$^{cut}_e$) & 10 GeV\\
                 & & Luminosity (L$_e$) & 6.7 $\times$ 10$^{32}$ erg/s\\
                 & & Magnetic field (B) & 0.04 G\\
                 & & Photon energy density (u$_{ph}$) & 0.20 erg cm$^{-3}$\\
                 & & Electron-to-proton luminosity ratio (L$_e$/L$_p$) & 7.3 $\times$ 10$^{-4}$\\
\enddata
\end{deluxetable*}

\begin{acknowledgments}
We thank the anonymous reviewer for helpful comments and constructive criticism. ADS thanks Nayantara Gupta for helpful discussions. SR acknowledges support from the University of Johannesburg Research Council and the National Institute for Theoretical Computational Sciences (NITheCS), South Africa. Nayana A.J. would like to acknowledge DST-INSPIRE Faculty Fellowship (IFA20-PH-259) for supporting this research.  
\end{acknowledgments}

\vspace{5mm}


\software{\texttt{GAMERA} (\url{https://github.com/libgamera/GAMERA})
          }



\pagebreak



\bibliography{sample631}{}

\begin{thebibliography}{}
\expandafter\ifx\csname natexlab\endcsname\relax\def\natexlab#1{#1}\fi
\providecommand{\url}[1]{\href{#1}{#1}}
\providecommand{\dodoi}[1]{doi:~\href{http://doi.org/#1}{\nolinkurl{#1}}}
\providecommand{\doeprint}[1]{\href{http://ascl.net/#1}{\nolinkurl{http://ascl.net/#1}}}
\providecommand{\doarXiv}[1]{\href{https://arxiv.org/abs/#1}{\nolinkurl{https://arxiv.org/abs/#1}}}

\bibitem[{{Abdo} {et~al.}(2010){Abdo}, {Ackermann}, {Ajello}, {Atwood},
  {Baldini}, {Ballet}, {Barbiellini}, {Bastieri}, {Bechtol}, {Bellazzini},
  {Berenji}, {Blandford}, \& {Fermi LAT Collaboration}}]{abdo10}
{Abdo}, A.~A., {Ackermann}, M., {Ajello}, M., {et~al.} 2010, Science, 329, 817,
  \dodoi{10.1126/science.1192537}

\bibitem[{{Acciari} {et~al.}(2022){Acciari}, {Ansoldi}, {Antonelli}, {Arbet
  Engels}, {Artero}, {Asano}, {Baack}, {Babi{\'c}}, {Baquero}, {Barres de
  Almeida}, {Barrio}, {Batkovi{\'c}}, {Becerra Gonz{\'a}lez}, {Bednarek},
  {Bellizzi}, {Bernardini}, {Bernardos}, {Berti}, {Besenrieder},
  {Bhattacharyya}, {Bigongiari}, {Biland}, {Blanch}, {B{\"o}kenkamp},
  {Bonnoli}, {Bo{\v{s}}njak}, {Busetto}, {Carosi}, {Ceribella}, {Cerruti},
  {Chai}, {Chilingarian}, {Cikota}, {Colak}, {Colombo}, {Contreras}, {Cortina},
  {Covino}, {D'Amico}, {D'Elia}, {Da Vela}, {Dazzi}, {De Angelis}, {De Lotto},
  {Del Popolo}, {Delfino}, {Delgado}, {Delgado Mendez}, {Depaoli}, {Di Pierro},
  {Di Venere}, {Do Souto Espi{\~n}eira}, {Prester}, {Donini}, {Dorner}, {Doro},
  {Elsaesser}, {Fallah Ramazani}, {Fari{\~n}a Alonso}, {Fattorini}, {Fonseca},
  {Font}, {Fruck}, {Fukami}, {Fukazawa}, {Garc{\'\i}a L{\'o}pez},
  {Garczarczyk}, {Gasparyan}, {Gaug}, {Giglietto}, {Giordano}, {Gliwny},
  {Godinovi{\'c}}, {Green}, {Green}, {Hadasch}, {Hahn}, {Hassan}, {Heckmann},
  {Herrera}, {Hoang}, {Hrupec}, {H{\"u}tten}, {Inada}, {Ishio}, {Iwamura},
  {Jim{\'e}nez Mart{\'\i}nez}, {Jormanainen}, {Jouvin}, {Kerszberg},
  {Kobayashi}, {Kubo}, {Kushida}, {Lamastra}, {Lelas}, {Leone}, {Lindfors},
  {Linhoff}, {Lombardi}, {Longo}, {L{\'o}pez-Coto}, {L{\'o}pez-Moya},
  {L{\'o}pez-Oramas}, {Loporchio}, {Machado de Oliveira Fraga}, {Maggio},
  {Majumdar}, {Makariev}, {Mallamaci}, {Maneva}, {Manganaro}, {Mannheim},
  {Maraschi}, {Mariotti}, {Mart{\'\i}nez}, {Mas Aguilar}, {Mazin}, {Menchiari},
  {Mender}, {Mi{\'c}anovi{\'c}}, {Miceli}, {Miener}, {Miranda}, {Mirzoyan},
  {Molina}, {Moralejo}, {Morcuende}, {Moreno}, {Moretti}, {Nakamori}, {Nava},
  {Neustroev}, {Nievas Rosillo}, {Nigro}, {Nilsson}, {Nishijima}, {Noda},
  {Nozaki}, {Ohtani}, {Oka}, {Otero-Santos}, {Paiano}, {Palatiello}, {Paneque},
  {Paoletti}, {Paredes}, {Pavleti{\'c}}, {Pe{\~n}il}, {Persic}, {Pihet}, {Prada
  Moroni}, {Prandini}, {Priyadarshi}, {Puljak}, {Rhode}, {Rib{\'o}}, {Rico},
  {Righi}, {Rugliancich}, {Sahakyan}, {Saito}, {Sakurai}, {Satalecka},
  {Saturni}, {Schleicher}, {Schmidt}, {Schweizer}, {Sitarek},
  {{\v{S}}nidari{\'c}}, {Sobczynska}, {Spolon}, {Stamerra},
  {Stri{\v{s}}kovi{\'c}}, {Strom}, {Strzys}, {Suda}, {Suri{\'c}}, {Takahashi},
  {Takeishi}, {Tavecchio}, {Temnikov}, {Terzi{\'c}}, {Teshima}, {Tosti},
  {Truzzi}, {Tutone}, {Ubach}, {van Scherpenberg}, {Vanzo}, {Vazquez Acosta},
  {Ventura}, {Verguilov}, {Vigorito}, {Vitale}, {Vovk}, {Will}, {Wunderlich},
  {Yamamoto}, {Zari{\'c}}, {Ambrosino}, {Cecconi}, {Catanzaro}, {Ferrara},
  {Frasca}, {Munari}, {Giustolisi}, {Alonso-Santiago}, {Giarrusso}, {Munari},
  \& {Valisa}}]{magic22}
{Acciari}, V.~A., {Ansoldi}, S., {Antonelli}, L.~A., {et~al.} 2022, Nature
  Astronomy, 6, 689, \dodoi{10.1038/s41550-022-01640-z}

\bibitem[{{Ackermann} {et~al.}(2014){Ackermann}, {Ajello}, {Albert}, {Baldini},
  {Ballet}, {Barbiellini}, {Bastieri}, {Bellazzini}, {Bissaldi}, {Blandford},
  {Bloom}, {Bottacini}, {Brandt}, {Bregeon}, {Bruel}, {Buehler}, {Buson},
  {Caliandro}, {Cameron}, {Caragiulo}, {Caraveo}, {Cavazzuti}, {Charles},
  {Chekhtman}, {Cheung}, {Chiang}, {Chiaro}, {Ciprini}, {Claus},
  {Cohen-Tanugi}, {Conrad}, {Corbel}, {D'Ammando}, {de Angelis}, {den Hartog},
  {de Palma}, {Dermer}, {Desiante}, {Digel}, {Di Venere}, {do Couto e Silva},
  {Donato}, {Drell}, {Drlica-Wagner}, {Favuzzi}, {Ferrara}, {Focke},
  {Franckowiak}, {Fuhrmann}, {Fukazawa}, {Fusco}, {Gargano}, {Gasparrini},
  {Germani}, {Giglietto}, {Giordano}, {Giroletti}, {Glanzman}, {Godfrey},
  {Grenier}, {Grove}, {Guiriec}, {Hadasch}, {Harding}, {Hayashida}, {Hays},
  {Hewitt}, {Hill}, {Hou}, {Jean}, {Jogler}, {J{\'o}hannesson}, {Johnson},
  {Johnson}, {Kerr}, {Kn{\"o}dlseder}, {Kuss}, {Larsson}, {Latronico},
  {Lemoine-Goumard}, {Longo}, {Loparco}, {Lott}, {Lovellette}, {Lubrano},
  {Manfreda}, {Martin}, {Massaro}, {Mayer}, {Mazziotta}, {McEnery},
  {Michelson}, {Mitthumsiri}, {Mizuno}, {Monzani}, {Morselli}, {Moskalenko},
  {Murgia}, {Nemmen}, {Nuss}, {Ohsugi}, {Omodei}, {Orienti}, {Orlando},
  {Ormes}, {Paneque}, {Panetta}, {Perkins}, {Pesce-Rollins}, {Piron}, {Pivato},
  {Porter}, {Rain{\`o}}, {Rando}, {Razzano}, {Razzaque}, {Reimer}, {Reimer},
  {Reposeur}, {Saz Parkinson}, {Schaal}, {Schulz}, {Sgr{\`o}}, {Siskind},
  {Spandre}, {Spinelli}, {Stawarz}, {Suson}, {Takahashi}, {Tanaka}, {Thayer},
  {Thayer}, {Thompson}, {Tibaldo}, {Tinivella}, {Torres}, {Tosti}, {Troja},
  {Uchiyama}, {Vianello}, {Winer}, {Wolff}, {Wood}, {Wood}, {Wood},
  {Charbonnel}, {Corbet}, {De Gennaro Aquino}, {Edlin}, {Mason}, {Schwarz},
  {Shore}, {Starrfield}, {Teyssier}, \& {Fermi-LAT
  Collaboration}}]{ackermann14}
{Ackermann}, M., {Ajello}, M., {Albert}, A., {et~al.} 2014, Science, 345, 554,
  \dodoi{10.1126/science.1253947}

\bibitem[{{Aharonian} {et~al.}(2022){Aharonian}, {Ait Benkhali}, {Ang{\"u}ner},
  {Ashkar}, {Backes}, {Baghmanyan}, {Barbosa Martins}, {Batzofin}, {Becherini},
  {Berge}, {Bernl{\"o}hr}, {Bi}, {B{\"o}ttcher}, {Boisson}, {Bolmont}, {de Bony
  de Lavergne}, {Breuhaus}, {Brose}, {Brun}, {Caroff}, {Casanova}, {Cerruti},
  {Chand}, {Chen}, {Cotter}, {Damascene Mbarubucyeye}, {Djannati-Ata{\"\i}},
  {Dmytriiev}, {Doroshenko}, {Duffy}, {Egberts}, {Ernenwein}, {Fegan},
  {Feijen}, {Fiasson}, {Fichet de Clairfontaine}, {Fontaine},
  {F{\"u}{\ss}ling}, {Funk}, {Gabici}, {Gallant}, {Ghafourizadeh}, {Giavitto},
  {Giunti}, {Glawion}, {Glicenstein}, {Grondin}, {Hermann}, {Hinton},
  {H{\"o}rbe}, {Hofmann}, {Hoischen}, {Holch}, {Holler}, {Horns}, {Huang},
  {Jamrozy}, {Jankowsky}, {Jung-Richardt}, {Kasai}, {Katarzy{\'n}ski}, {Katz},
  {Khangulyan}, {Kh{\'e}lifi}, {Klepser}, {Klu{\'z}niak}, {Komin}, {Konno},
  {Kosack}, {Kostunin}, {Le Stum}, {Lemi{\`e}re}, {Lemoine-Goumard}, {Lenain},
  {Leuschner}, {Lohse}, {Luashvili}, {Lypova}, {Mackey}, {Malyshev},
  {Malyshev}, {Marandon}, {Marchegiani}, {Marcowith}, {Mart{\'\i}-Devesa},
  {Marx}, {Maurin}, {Meyer}, {Mitchell}, {Moderski}, {Mohrmann}, {Montanari},
  {Moulin}, {Muller}, {Murach}, {Nakashima}, {de Naurois}, {Nayerhoda},
  {Niemiec}, {Priyana Noel}, {O{\textquoteright}Brien}, {Ohm}, {Olivera-Nieto},
  {de Ona Wilhelmi}, {Ostrowski}, {Panny}, {Panter}, {Parsons}, {Peron},
  {Pita}, {Poireau}, {Prokhorov}, {Prokoph}, {P{\"u}hlhofer}, {Punch},
  {Quirrenbach}, {Reichherzer}, {Reimer}, {Reimer}, {Renaud}, {Reville},
  {Rieger}, {Rowell}, {Rudak}, {Rueda Ricarte}, {Ruiz-Velasco}, {Sahakian},
  {Sailer}, {Salzmann}, {Sanchez}, {Santangelo}, {Sasaki}, {Sch{\"a}fer},
  {Sch{\"u}ssler}, {Schutte}, {Schwanke}, {Senniappan}, {Shapopi}, {Simoni},
  {Sinha}, {Sol}, {Specovius}, {Spencer}, {Stawarz}, {Steinmassl}, {Steppa},
  {Takahashi}, {Tanaka}, {Taylor}, {Terrier}, {Thorpe-Morgan}, {Tsirou},
  {Tsuji}, {Tuffs}, {Uchiyama}, {Unbehaun}, {van Eldik}, {van Soelen}, {Veh},
  {Venter}, {Vink}, {Wagner}, {Werner}, {White}, {Wierzcholska}, {Wong},
  {Yusafzai}, {Zacharias}, {Zargaryan}, {Zdziarski}, {Zech}, {Zhu}, {Zouari},
  \& {{\.Z}ywucka}}]{hess22}
{Aharonian}, F., {Ait Benkhali}, F., {Ang{\"u}ner}, E.~O., {et~al.} 2022,
  Science, 376, 77, \dodoi{10.1126/science.abn0567}

\bibitem[{{Ahnen} {et~al.}(2015){Ahnen}, {Ansoldi}, {Antonelli}, {Antoranz},
  {Babic}, {Banerjee}, {Bangale}, {Barres de Almeida}, {Barrio}, {Becerra
  Gonz{\'a}lez}, {Bednarek}, {Bernardini}, {Biasuzzi}, {Biland}, {Blanch},
  {Bonnefoy}, {Bonnoli}, {Borracci}, {Bretz}, {Carmona}, {Carosi},
  {Chatterjee}, {Clavero}, {Colin}, {Colombo}, {Contreras}, {Cortina},
  {Covino}, {Da Vela}, {Dazzi}, {De Angelis}, {De Caneva}, {De Lotto}, {de
  O{\~n}a Wilhelmi}, {Delgado Mendez}, {Di Pierro}, {Dominis Prester},
  {Dorner}, {Doro}, {Einecke}, {Eisenacher Glawion}, {Elsaesser},
  {Fern{\'a}ndez-Barral}, {Fidalgo}, {Fonseca}, {Font}, {Frantzen}, {Fruck},
  {Galindo}, {Garc{\'\i}a L{\'o}pez}, {Garczarczyk}, {Garrido Terrats}, {Gaug},
  {Giammaria}, {Godinovi{\'c}}, {Gonz{\'a}lez Mu{\~n}oz}, {Guberman},
  {Hanabata}, {Hayashida}, {Herrera}, {Hose}, {Hrupec}, {Hughes}, {Idec},
  {Kellermann}, {Kodani}, {Konno}, {Kubo}, {Kushida}, {La Barbera}, {Lelas},
  {Lewandowska}, {Lindfors}, {Lombardi}, {Longo}, {L{\'o}pez},
  {L{\'o}pez-Coto}, {L{\'o}pez-Oramas}, {Lorenz}, {Majumdar}, {Makariev},
  {Mallot}, {Maneva}, {Manganaro}, {Mannheim}, {Maraschi}, {Marcote},
  {Mariotti}, {Mart{\'\i}nez}, {Mazin}, {Menzel}, {Miranda}, {Mirzoyan},
  {Moralejo}, {Nakajima}, {Neustroev}, {Niedzwiecki}, {Nievas Rosillo},
  {Nilsson}, {Nishijima}, {Noda}, {Orito}, {Overkemping}, {Paiano}, {Palacio},
  {Palatiello}, {Paneque}, {Paoletti}, {Paredes}, {Paredes-Fortuny}, {Persic},
  {Poutanen}, {Prada Moroni}, {Prandini}, {Puljak}, {Reinthal}, {Rhode},
  {Rib{\'o}}, {Rico}, {Rodriguez Garcia}, {Saito}, {Saito}, {Satalecka},
  {Scapin}, {Schultz}, {Schweizer}, {Sillanp{\"a}{\"a}}, {Sitarek}, {Snidaric},
  {Sobczynska}, {Stamerra}, {Steinbring}, {Strzys}, {Takalo}, {Takami},
  {Tavecchio}, {Temnikov}, {Terzi{\'c}}, {Tescaro}, {Teshima}, {Thaele},
  {Torres}, {Toyama}, {Treves}, {Verguilov}, {Vovk}, {Will}, {Zanin},
  {Desiante}, \& {Hays}}]{magic15}
{Ahnen}, M.~L., {Ansoldi}, S., {Antonelli}, L.~A., {et~al.} 2015, \aap, 582,
  A67, \dodoi{10.1051/0004-6361/201526478}

\bibitem[{{Aiello} {et~al.}(2019){Aiello}, {Akrame}, {Ameli}, {Anassontzis},
  {Andre}, {Androulakis}, {Anghinolfi}, {Anton}, {Ardid}, {Aublin}, {Avgitas},
  {Bagatelas}, {Barbarino}, {Baret}, {Barrios-Mart{\'\i}}, {Belias}, {Berbee},
  {van den Berg}, {Bertin}, {Biagi}, {Biagioni}, {Biernoth}, {Boumaaza},
  {Bourret}, {Bouta}, {Bouwhuis}, {Bozza}, {Br{\^a}nza{\c{s}}}, {Bruchner},
  {Bruijn}, {Brunner}, {Buis}, {Buompane}, {Busto}, {Calvo}, {Capone}, {Celli},
  {Chabab}, {Chau}, {Cherubini}, {Chiarella}, {Chiarusi}, {Circella},
  {Cocimano}, {Coelho}, {Coleiro}, {Molla}, {Coniglione}, {Coyle}, {Creusot},
  {Cuttone}, {D'Onofrio}, {Dallier}, {De Sio}, {Di Palma}, {D{\'\i}az},
  {Diego-Tortosa}, {Distefano}, {Domi}, {Don{\`a}}, {Donzaud}, {Dornic},
  {D{\"o}rr}, {Durocher}, {Eberl}, {van Eijk}, {El Bojaddaini}, {Eljarrari},
  {Elsaesser}, {Enzenh{\"o}fer}, {Fermani}, {Ferrara}, {Filipovi{\'c}},
  {Fusco}, {Gal}, {Garcia}, {Garufi}, {Gialanella}, {Giorgio}, {Giuliante},
  {Gozzini}, {Gracia}, {Graf}, {Grasso}, {Gr{\'e}goire}, {Grella}, {Hallmann},
  {Hamdaoui}, {van Haren}, {Heid}, {Heijboer}, {Hekalo}, {Hern{\'a}ndez-Rey},
  {Hofest{\"a}dt}, {Illuminati}, {James}, {Jongen}, {de Jong}, {de Jong},
  {Kadler}, {Kalaczy{\'n}ski}, {Kalekin}, {Katz}, {Khan Chowdhury},
  {Kie{\ss}ling}, {Koffeman}, {Kooijman}, {Kouchner}, {Kreter}, {Kulikovskiy},
  {Kunhikannan Kannichankandy}, {Lahmann}, {Larosa}, {Le Breton}, {Leone},
  {Leonora}, {Levi}, {Lincetto}, {Lonardo}, {Longhitano}, {Lopez Coto},
  {Lotze}, {Maderer}, {Maggi}, {Ma{\'n}czak}, {Mannheim}, {Margiotta},
  {Marinelli}, {Markou}, {Martin}, {Mart{\'\i}nez-Mora}, {Martini},
  {Marzaioli}, {Mele}, {Melis}, {Migliozzi}, {Migneco}, {Mijakowski},
  {Miranda}, {Mollo}, {Morganti}, {Moser}, {Moussa}, {Muller}, {Musumeci},
  {Nauta}, {Navas}, {Nicolau}, {Nielsen}, {{\'O} Fearraigh}, {Organokov},
  {Orlando}, {Ottonello}, {Panagopoulos}, {Papalashvili}, {Papaleo},
  {P{\u{a}}v{\u{a}}la{\c{s}}}, {Pellegrino}, {Perrin-Terrin}, {Piattelli},
  {Pikounis}, {Pisanti}, {Poir{\'e}}, {Polydefki}, {Popa}, {Post}, {Pradier},
  {P{\"u}hlhofer}, {Pulvirenti}, {Quinn}, {Raffaelli}, {Randazzo}, {Razzaque},
  {Real}, {Resvanis}, {Reubelt}, {Riccobene}, {Richer}, {Rigalleau}, {Rovelli},
  {Saffer}, {Salvadori}, {Samtleben}, {S{\'a}nchez Losa}, {Sanguineti},
  {Santangelo}, {Santonocito}, {Sapienza}, {Schumann}, {Sciacca}, {Seneca},
  {Sgura}, {Shanidze}, {Sharma}, {Simeone}, {Sinopoulou}, {Spisso}, {Spurio},
  {Stavropoulos}, {Steijger}, {Stellacci}, {Strandberg}, {Stransky},
  {St{\"u}ven}, {Taiuti}, {Tatone}, {Tayalati}, {Tenllado}, {Thakore},
  {Trovato}, {Tzamariudaki}, {Tzanetatos}, {Van Elewyck}, {Versari}, {Viola},
  {Vivolo}, {Wilms}, {de Wolf}, {Zaborov}, {Zornoza}, {Z{\'u}{\~n}iga}, \&
  {KM3NeT Collaboration}}]{aiello19}
{Aiello}, S., {Akrame}, S.~E., {Ameli}, F., {et~al.} 2019, Astroparticle
  Physics, 111, 100, \dodoi{10.1016/j.astropartphys.2019.04.002}

\bibitem[{{Albert} {et~al.}(2019){Albert}, {Alfaro}, {Ashkar}, {Alvarez},
  {{\'A}lvarez}, {Arteaga-Vel{\'a}zquez}, {Ayala Solares}, {Arceo}, {Bellido},
  {BenZvi}, {Bretz}, {Brisbois}, {Brown}, {Brun}, {Caballero-Mora}, {Carosi},
  {Carrami{\~n}ana}, {Casanova}, {Chadwick}, {Cotter}, {Couti{\~n}o De
  Le{\'o}n}, {Cristofari}, {Dasso}, {de la Fuente}, {Dingus}, {Desiati},
  {Salles}, {de Souza}, {Dorner}, {D{\'\i}az-V{\'e}lez},
  {Garc{\'\i}a-Gonz{\'a}lez}, {DuVernois}, {Di Sciascio}, {Engel},
  {Fleischhack}, {Fraija}, {Funk}, {Glicenstein}, {Gonzalez}, {Gonz{\'a}lez},
  {Goodman}, {Harding}, {Haungs}, {Hinton}, {Hona}, {Hoyos}, {Huentemeyer},
  {Iriarte}, {Jardin-Blicq}, {Joshi}, {Kaufmann}, {Kawata}, {Kunwar},
  {Lefaucheur}, {Lenain}, {Link}, {L{\'o}pez-Coto}, {Marandon}, {Mariotti},
  {Mart{\'\i}nez-Castro}, {Mart{\'\i}nez-Huerta}, {Mostaf{\'a}}, {Nayerhoda},
  {Nellen}, {de O{\~n}a Wilhelmi}, {Parsons}, {Patricelli}, {Pichel}, {Piel},
  {Prandini}, {Pueschel}, {Procureur}, {Reisenegger}, {Rivi{\`e}re},
  {Rodriguez}, {Rovero}, {Rowell}, {Ruiz-Velasco}, {Sandoval}, {Santander},
  {Sako}, {Sako}, {Satalecka}, {Schoorlemmer}, {Sch{\"u}ssler},
  {Seglar-Arroyo}, {Smith}, {Spencer}, {Surajbali}, {Tabachnick}, {Taylor},
  {Tibolla}, {Torres}, {Vallage}, {Viana}, {Watson}, {Weisgarber}, {Werner},
  {White}, {Wischnewski}, {Yang}, {Zepeda}, \& {Zhou}}]{SWGO19}
{Albert}, A., {Alfaro}, R., {Ashkar}, H., {et~al.} 2019, arXiv e-prints,
  arXiv:1902.08429.
\newblock \doarXiv{1902.08429}

\bibitem[{{Anupama} \& {Miko{\l}ajewska}(1999)}]{anupama99}
{Anupama}, G.~C., \& {Miko{\l}ajewska}, J. 1999, \aap, 344, 177,
  \dodoi{10.48550/arXiv.astro-ph/9812432}

\bibitem[{{Anupama} {et~al.}(2013){Anupama}, {Kamath}, {Ramaprakash},
  {Kantharia}, {Hegde}, {Mohan}, {Kulkarni}, {Bode}, {Eyres}, {Evans}, \&
  {O'Brien}}]{anupama13}
{Anupama}, G.~C., {Kamath}, U.~S., {Ramaprakash}, A.~N., {et~al.} 2013, \aap,
  559, A121, \dodoi{10.1051/0004-6361/201321262}

\bibitem[{{Baring} {et~al.}(1999){Baring}, {Ellison}, {Reynolds}, {Grenier}, \&
  {Goret}}]{baring}
{Baring}, M.~G., {Ellison}, D.~C., {Reynolds}, S.~P., {Grenier}, I.~A., \&
  {Goret}, P. 1999, The Astrophysical Journal, 513, 311, \dodoi{10.1086/306829}

\bibitem[{{Barry} {et~al.}(2008){Barry}, {Mukai}, {Sokoloski}, {Danchi},
  {Hachisu}, {Evans}, {Gehrz}, \& {Mikolajewska}}]{barry08}
{Barry}, R.~K., {Mukai}, K., {Sokoloski}, J.~L., {et~al.} 2008, in Astronomical
  Society of the Pacific Conference Series, Vol. 401, RS Ophiuchi (2006) and
  the Recurrent Nova Phenomenon, ed. A.~{Evans}, M.~F. {Bode}, T.~J. {O'Brien},
  \& M.~J. {Darnley}, 52

\bibitem[{{Barry} {et~al.}(2007){Barry}, {Danchi}, {Sokoloski}, {Koresko},
  {Wisniewski}, {Serabyn}, {Traub}, {Kuchner}, \& {Greenhouse}}]{barry07}
{Barry}, R.~K., {Danchi}, W.~C., {Sokoloski}, J.~L., {et~al.} 2007, in American
  Astronomical Society Meeting Abstracts, Vol. 211, American Astronomical
  Society Meeting Abstracts, 57.05

\bibitem[{{Bednarek} \& {Pabich}(2011)}]{bednarek11}
{Bednarek}, W., \& {Pabich}, J. 2011, \aap, 530, A49,
  \dodoi{10.1051/0004-6361/201116549}

\bibitem[{{Blumenthal} \& {Gould}(1970)}]{blumenthal}
{Blumenthal}, G.~R., \& {Gould}, R.~J. 1970, Reviews of Modern Physics, 42,
  237, \dodoi{10.1103/RevModPhys.42.237}

\bibitem[{{Bode} \& {Evans}(2008)}]{bode08}
{Bode}, M.~F., \& {Evans}, A. 2008, {Classical Novae}, Vol.~43

\bibitem[{{Bode} \& {Kahn}(1985)}]{bode85}
{Bode}, M.~F., \& {Kahn}, F.~D. 1985, \mnras, 217, 205,
  \dodoi{10.1093/mnras/217.1.205}

\bibitem[{{Booth} {et~al.}(2016){Booth}, {Mohamed}, \&
  {Podsiadlowski}}]{booth16}
{Booth}, R.~A., {Mohamed}, S., \& {Podsiadlowski}, P. 2016, \mnras, 457, 822,
  \dodoi{10.1093/mnras/stw001}

\bibitem[{{Brandi} {et~al.}(2009){Brandi}, {Quiroga}, {Miko{\l}ajewska},
  {Ferrer}, \& {Garc{\'\i}a}}]{brandi09}
{Brandi}, E., {Quiroga}, C., {Miko{\l}ajewska}, J., {Ferrer}, O.~E., \&
  {Garc{\'\i}a}, L.~G. 2009, \aap, 497, 815,
  \dodoi{10.1051/0004-6361/200811417}

\bibitem[{{Cherenkov Telescope Array Consortium} {et~al.}(2019){Cherenkov
  Telescope Array Consortium}, {Acharya}, {Agudo}, {Al Samarai}, {Alfaro},
  {Alfaro}, {Alispach}, {Alves Batista}, {Amans}, {Amato}, {Ambrosi},
  {Antolini}, {Antonelli}, {Aramo}, {Araya}, {Armstrong}, {Arqueros},
  {Arrabito}, {Asano}, {Ashley}, {Backes}, {Balazs}, {Balbo}, {Ballester},
  {Ballet}, {Bamba}, {Barkov}, {Barres de Almeida}, {Barrio}, {Bastieri},
  {Becherini}, {Belfiore}, {Benbow}, {Berge}, {Bernardini}, {Bernardini},
  {Bernardos}, {Bernl{\"o}hr}, {Bertucci}, {Biasuzzi}, {Bigongiari}, {Biland},
  {Bissaldi}, {Biteau}, {Blanch}, {Blazek}, {Boisson}, {Bolmont}, {Bonanno},
  {Bonardi}, {Bonavolont{\`a}}, {Bonnoli}, {Bosnjak}, {B{\"o}ttcher},
  {Braiding}, {Bregeon}, {Brill}, {Brown}, {Brun}, {Brunetti}, {Buanes},
  {Buckley}, {Bugaev}, {B{\"u}hler}, {Bulgarelli}, {Bulik}, {Burton},
  {Burtovoi}, {Busetto}, {Canestrari}, {Capalbi}, {Capitanio}, {Caproni},
  {Caraveo}, {C{\'a}rdenas}, {Carlile}, {Carosi}, {Carqu{\'\i}n}, {Carr},
  {Casanova}, {Cascone}, {Catalani}, {Catalano}, {Cauz}, {Cerruti}, {Chadwick},
  {Chaty}, {Chaves}, {Chen}, {Chen}, {Chernyakova}, {Chikawa}, {Christov},
  {Chudoba}, {Cie{\'s}lar}, {Coco}, {Colafrancesco}, {Colin}, {Conforti},
  {Connaughton}, {Conrad}, {Contreras}, {Cortina}, {Costa}, {Costantini},
  {Cotter}, {Covino}, {Crocker}, {Cuadra}, {Cuevas}, {Cumani}, {D'A{\`\i}},
  {D'Ammando}, {D'Avanzo}, {D'Urso}, {Daniel}, {Davids}, {Dawson}, {Dazzi}, {De
  Angelis}, {de C{\'a}ssia dos Anjos}, {De Cesare}, {De Franco}, {de Gouveia
  Dal Pino}, {de la Calle}, {de los Reyes Lopez}, {De Lotto}, {De Luca}, {De
  Lucia}, {de Naurois}, {de O{\~n}a Wilhelmi}, {De Palma}, {De Persio}, {de
  Souza}, {Deil}, {Del Santo}, {Delgado}, {della Volpe}, {Di Girolamo}, {Di
  Pierro}, {Di Venere}, {D{\'\i}az}, {Dib}, {Diebold}, {Djannati-Ata{\"\i}},
  {Dom{\'\i}nguez}, {Dominis Prester}, {Dorner}, {Doro}, {Drass}, {Dravins},
  {Dubus}, {Dwarkadas}, {Ebr}, {Eckner}, {Egberts}, {Einecke}, {Ekoume},
  {Els{\"a}sser}, {Ernenwein}, {Espinoza}, {Evoli}, {Fairbairn},
  {Falceta-Goncalves}, {Falcone}, {Farnier}, {Fasola}, {Fedorova}, {Fegan},
  {Fernandez-Alonso}, {Fern{\'a}ndez-Barral}, {Ferrand}, {Fesquet},
  {Filipovic}, {Fioretti}, {Fontaine}, {Fornasa}, {Fortson}, {Freixas
  Coromina}, {Fruck}, {Fujita}, {Fukazawa}, {Funk}, {F{\"u}{\ss}ling},
  {Gabici}, {Gadola}, {Gallant}, {Garcia}, {Garcia L{\'o}pez}, {Garczarczyk},
  {Gaskins}, {Gasparetto}, {Gaug}, {Gerard}, {Giavitto}, {Giglietto}, {Giommi},
  {Giordano}, {Giro}, {Giroletti}, {Giuliani}, {Glicenstein}, {Gnatyk},
  {Godinovic}, {Goldoni}, {G{\'o}mez-Vargas}, {Gonz{\'a}lez}, {Gonz{\'a}lez},
  {G{\"o}tz}, {Graham}, {Grandi}, {Granot}, {Green}, {Greenshaw}, {Griffiths},
  {Gunji}, {Hadasch}, {Hara}, {Hardcastle}, {Hassan}, {Hayashi}, {Hayashida},
  {Heller}, {Helo}, {Hermann}, {Hinton}, {Hnatyk}, {Hofmann}, {Holder},
  {Horan}, {H{\"o}randel}, {Horns}, {Horvath}, {Hovatta}, {Hrabovsky},
  {Hrupec}, {Humensky}, {H{\"u}tten}, {Iarlori}, {Inada}, {Inome}, {Inoue},
  {Inoue}, {Inoue}, {Iocco}, {Ioka}, {Iori}, {Ishio}, {Iwamura}, {Jamrozy},
  {Janecek}, {Jankowsky}, {Jean}, {Jung-Richardt}, {Jurysek}, {Kaaret},
  {Karkar}, {Katagiri}, {Katz}, {Kawanaka}, {Kazanas}, {Kh{\'e}lifi}, {Kieda},
  {Kimeswenger}, {Kimura}, {Kisaka}, {Knapp}, {Kn{\"o}dlseder}, {Koch},
  {Kohri}, {Komin}, {Kosack}, {Kraus}, {Krause}, {Krau{\ss}}, {Kubo}, {Kukec
  Mezek}, {Kuroda}, {Kushida}, {La Palombara}, {Lamanna}, {Lang}, {Lapington},
  {Le Blanc}, {Leach}, {Lees}, {Lefaucheur}, {Leigui de Oliveira}, {Lenain},
  {Lico}, {Limon}, {Lindfors}, {Lohse}, {Lombardi}, {Longo}, {L{\'o}pez},
  {L{\'o}pez-Coto}, {Lu}, {Lucarelli}, {Luque-Escamilla}, {Lyard}, {Maccarone},
  {Maier}, {Majumdar}, {Malaguti}, {Mandat}, {Maneva}, {Manganaro}, {Mangano},
  {Marcowith}, {Mar{\'\i}n}, {Markoff}, {Mart{\'\i}}, {Martin},
  {Mart{\'\i}nez}, {Mart{\'\i}nez}, {Masetti}, {Masuda}, {Maurin}, {Maxted},
  {Mazin}, {Medina}, {Melandri}, {Mereghetti}, {Meyer}, {Minaya}, {Mirabal},
  {Mirzoyan}, {Mitchell}, {Mizuno}, {Moderski}, {Mohammed}, {Mohrmann},
  {Montaruli}, {Moralejo}, {Morcuende-Parrilla}, {Mori}, {Morlino}, {Morris},
  {Morselli}, {Moulin}, {Mukherjee}, {Mundell}, {Murach}, {Muraishi}, {Murase},
  {Nagai}, {Nagataki}, {Nagayoshi}, {Naito}, {Nakamori}, {Nakamura}, {Niemiec},
  {Nieto}, {Niko{\l}ajuk}, {Nishijima}, {Noda}, {Nosek}, {Novosyadlyj},
  {Nozaki}, {O'Brien}, {Oakes}, {Ohira}, {Ohishi}, {Ohm}, {Okazaki}, {Okumura},
  {Ong}, {Orienti}, {Orito}, {Osborne}, {Ostrowski}, {Otte}, {Oya}, {Padovani},
  {Paizis}, {Palatiello}, {Palatka}, {Paoletti}, {Paredes}, {Pareschi},
  {Parsons}, {Pe'er}, {Pech}, {Pedaletti}, {Perri}, {Persic}, {Petrashyk},
  {Petrucci}, {Petruk}, {Peyaud}, {Pfeifer}, {Piano}, {Pisarski}, {Pita},
  {Pohl}, {Polo}, {Pozo}, {Prandini}, {Prast}, {Principe}, {Prokhorov},
  {Prokoph}, {Prouza}, {P{\"u}hlhofer}, {Punch}, {P{\"u}rckhauer}, {Queiroz},
  {Quirrenbach}, {Rain{\`o}}, {Razzaque}, {Reimer}, {Reimer}, {Reisenegger},
  {Renaud}, {Rezaeian}, {Rhode}, {Ribeiro}, {Rib{\'o}}, {Richtler}, {Rico},
  {Rieger}, {Riquelme}, {Rivoire}, {Rizi}, {Rodriguez}, {Rodriguez Fernandez},
  {Rodr{\'\i}guez V{\'a}zquez}, {Rojas}, {Romano}, {Romeo}, {Rosado}, {Rovero},
  {Rowell}, {Rudak}, {Rugliancich}, {Rulten}, {Sadeh}, {Safi-Harb}, {Saito},
  {Sakaki}, {Sakurai}, {Salina}, {S{\'a}nchez-Conde}, {Sandaker}, {Sandoval},
  {Sangiorgi}, {Sanguillon}, {Sano}, {Santander}, {Sarkar}, {Satalecka},
  {Saturni}, {Schioppa}, {Schlenstedt}, {Schneider}, {Schoorlemmer},
  {Schovanek}, {Schulz}, {Schussler}, {Schwanke}, {Sciacca}, {Scuderi},
  {Seitenzahl}, {Semikoz}, {Sergijenko}, {Servillat}, {Shalchi}, {Shellard},
  {Sidoli}, {Siejkowski}, {Sillanp{\"a}{\"a}}, {Sironi}, {Sitarek}, {Sliusar},
  {Slowikowska}, {Sol}, {Stamerra}, {Stani{\v{c}}}, {Starling}, {Stawarz},
  {Stefanik}, {Stephan}, {Stolarczyk}, {Stratta}, {Straumann}, {Suomijarvi},
  {Supanitsky}, {Tagliaferri}, {Tajima}, {Tavani}, {Tavecchio}, {Tavernet},
  {Tayabaly}, {Tejedor}, {Temnikov}, {Terada}, {Terrier}, {Terzic}, {Teshima},
  {Testa}, {Thoudam}, {Tian}, {Tibaldo}, {Tluczykont}, {Todero Peixoto},
  {Tokanai}, {Tomastik}, {Tonev}, {Tornikoski}, {Torres}, {Torresi}, {Tosti},
  {Tothill}, {Tovmassian}, {Travnicek}, {Trichard}, {Trifoglio}, {Troyano
  Pujadas}, {Tsujimoto}, {Umana}, {Vagelli}, {Vagnetti}, {Valentino},
  {Vallania}, {Valore}, {van Eldik}, {Vandenbroucke}, {Varner}, {Vasileiadis},
  {Vassiliev}, {V{\'a}zquez Acosta}, {Vecchi}, {Vega}, {Vercellone}, {Veres},
  {Vergani}, {Verzi}, {Vettolani}, {Viana}, {Vigorito}, {Villanueva}, {Voelk},
  {Vollhardt}, {Vorobiov}, {Vrastil}, {Vuillaume}, {Wagner}, {Wagner},
  {Walter}, {Ward}, {Warren}, {Watson}, {Werner}, {White}, {White},
  {Wierzcholska}, {Wilcox}, {Will}, {Williams}, {Wischnewski}, {Wood},
  {Yamamoto}, {Yamazaki}, {Yanagita}, {Yang}, {Yoshida}, {Yoshiike},
  {Yoshikoshi}, {Zacharias}, {Zaharijas}, {Zampieri}, {Zandanel}, {Zanin},
  {Zavrtanik}, {Zavrtanik}, {Zdziarski}, {Zech}, {Zechlin}, {Zhdanov},
  {Ziegler}, \& {Zorn}}]{CTA19}
{Cherenkov Telescope Array Consortium}, {Acharya}, B.~S., {Agudo}, I., {et~al.}
  2019, {Science with the Cherenkov Telescope Array}, \dodoi{10.1142/10986}

\bibitem[{Cheung {et~al.}(2016)Cheung, Jean, Shore, Stawarz, Corbet,
  Knödlseder, Starrfield, Wood, Desiante, Longo, Pivato, \& Wood}]{cheung16}
Cheung, C.~C., Jean, P., Shore, S.~N., {et~al.} 2016, The Astrophysical
  Journal, 826, 142, \dodoi{10.3847/0004-637x/826/2/142}

\bibitem[{{Cheung} {et~al.}(2022){Cheung}, {Johnson}, {Jean}, {Kerr}, {Page},
  {Osborne}, {Beardmore}, {Sokolovsky}, {Teyssier}, {Ciprini},
  {Mart{\'\i}-Devesa}, {Mereu}, {Razzaque}, {Wood}, {Shore}, {Korotkiy},
  {Levina}, \& {Blumenzweig}}]{fermi22}
{Cheung}, C.~C., {Johnson}, T.~J., {Jean}, P., {et~al.} 2022, \apj, 935, 44,
  \dodoi{10.3847/1538-4357/ac7eb7}

\bibitem[{{Chomiuk} {et~al.}(2021){Chomiuk}, {Metzger}, \& {Shen}}]{chomiuk21}
{Chomiuk}, L., {Metzger}, B.~D., \& {Shen}, K.~J. 2021, \araa, 59, 391,
  \dodoi{10.1146/annurev-astro-112420-114502}

\bibitem[{{Chomiuk} {et~al.}(2012){Chomiuk}, {Krauss}, {Rupen}, {Nelson},
  {Roy}, {Sokoloski}, {Mukai}, {Munari}, {Mioduszewski}, {Weston}, {O'Brien},
  {Eyres}, \& {Bode}}]{chomiuk12}
{Chomiuk}, L., {Krauss}, M.~I., {Rupen}, M.~P., {et~al.} 2012, \apj, 761, 173,
  \dodoi{10.1088/0004-637X/761/2/173}

\bibitem[{{de Ruiter} {et~al.}(2023){de Ruiter}, {Nyamai}, {Rowlinson},
  {Wijers}, {O'Brien}, {Williams}, \& {Woudt}}]{ruiter23}
{de Ruiter}, I., {Nyamai}, M.~M., {Rowlinson}, A., {et~al.} 2023, arXiv
  e-prints, arXiv:2301.10552, \dodoi{10.48550/arXiv.2301.10552}

\bibitem[{{Diesing} {et~al.}(2023){Diesing}, {Metzger}, {Aydi}, {Chomiuk},
  {Vurm}, {Gupta}, \& {Caprioli}}]{diesing22}
{Diesing}, R., {Metzger}, B.~D., {Aydi}, E., {et~al.} 2023, \apj, 947, 70,
  \dodoi{10.3847/1538-4357/acc105}

\bibitem[{{Dobrzycka} {et~al.}(1994){Dobrzycka}, {Kenyon}, \&
  {Mikolajewska}}]{dob94}
{Dobrzycka}, D., {Kenyon}, S.~J., \& {Mikolajewska}, J. 1994, in Astronomical
  Society of the Pacific Conference Series, Vol.~56, Interacting Binary Stars,
  ed. A.~W. {Shafter}, 368

\bibitem[{{Franckowiak} {et~al.}(2018){Franckowiak}, {Jean}, {Wood}, {Cheung},
  \& {Buson}}]{franck18}
{Franckowiak}, A., {Jean}, P., {Wood}, M., {Cheung}, C.~C., \& {Buson}, S.
  2018, \aap, 609, A120, \dodoi{10.1051/0004-6361/201731516}

\bibitem[{{Ghisellini} {et~al.}(1988){Ghisellini}, {Guilbert}, \&
  {Svensson}}]{ghisellini}
{Ghisellini}, G., {Guilbert}, P.~W., \& {Svensson}, R. 1988, The Astrophysical
  Journal: Letters, 334, L5, \dodoi{10.1086/185300}

\bibitem[{{Gomez-Gomar} {et~al.}(1998){Gomez-Gomar}, {Hernanz}, {Jose}, \&
  {Isern}}]{gomez98}
{Gomez-Gomar}, J., {Hernanz}, M., {Jose}, J., \& {Isern}, J. 1998, \mnras, 296,
  913, \dodoi{10.1046/j.1365-8711.1998.01421.x}

\bibitem[{Gordon {et~al.}(2021)Gordon, Aydi, Page, Li, Chomiuk, Sokolovsky,
  Mukai, \& Seitz}]{gordon21}
Gordon, A.~C., Aydi, E., Page, K.~L., {et~al.} 2021, The Astrophysical Journal,
  910, 134, \dodoi{10.3847/1538-4357/abe547}

\bibitem[{{Guetta} {et~al.}(2023){Guetta}, {Hillman}, \& {Della
  Valle}}]{guetta22}
{Guetta}, D., {Hillman}, Y., \& {Della Valle}, M. 2023, \jcap, 2023, 015,
  \dodoi{10.1088/1475-7516/2023/03/015}

\bibitem[{Hahn(2016)}]{hahn}
Hahn, J. 2016, PoS, ICRC2015, 917, \dodoi{10.22323/1.236.0917}

\bibitem[{{Hellier}(2001)}]{hellier01}
{Hellier}, C. 2001, {Cataclysmic Variable Stars}

\bibitem[{{Kafexhiu} {et~al.}(2014){Kafexhiu}, {Aharonian}, {Taylor}, \&
  {Vila}}]{kafe}
{Kafexhiu}, E., {Aharonian}, F., {Taylor}, A.~M., \& {Vila}, G.~S. 2014,
  Physical Review D, 90, 123014, \dodoi{10.1103/PhysRevD.90.123014}

\bibitem[{{Kafka}(2021)}]{kafka}
{Kafka}, S. 2021, Observations from the AAVSO International Database

\bibitem[{{Kantharia} {et~al.}(2007){Kantharia}, {Anupama}, {Prabhu}, {Ramya},
  {Bode}, {Eyres}, \& {O'Brien}}]{kantharia07}
{Kantharia}, N.~G., {Anupama}, G.~C., {Prabhu}, T.~P., {et~al.} 2007, \apjl,
  667, L171, \dodoi{10.1086/522201}

\bibitem[{{Kantharia} {et~al.}(2016){Kantharia}, {Dutta}, {Roy}, {Anupama},
  {Ishwara-Chandra}, {Chitale}, {Prabhu}, {Banerjee}, \& {Ashok}}]{kantharia16}
{Kantharia}, N.~G., {Dutta}, P., {Roy}, N., {et~al.} 2016, \mnras, 456, L49,
  \dodoi{10.1093/mnrasl/slv154}

\bibitem[{{Kelner} {et~al.}(2006){Kelner}, {Aharonian}, \& {Bugayov}}]{kelner}
{Kelner}, S.~R., {Aharonian}, F.~A., \& {Bugayov}, V.~V. 2006, Physical Review
  D, 74, 034018, \dodoi{10.1103/PhysRevD.74.034018}

\bibitem[{{Knigge} {et~al.}(2011){Knigge}, {Baraffe}, \&
  {Patterson}}]{knigge11}
{Knigge}, C., {Baraffe}, I., \& {Patterson}, J. 2011, \apjs, 194, 28,
  \dodoi{10.1088/0067-0049/194/2/28}

\bibitem[{{Martin} \& {Dubus}(2013)}]{martin13}
{Martin}, P., \& {Dubus}, G. 2013, \aap, 551, A37,
  \dodoi{10.1051/0004-6361/201220289}

\bibitem[{{Miko{\l}ajewska}(2012)}]{miko12}
{Miko{\l}ajewska}, J. 2012, Baltic Astronomy, 21, 5,
  \dodoi{10.1515/astro-2017-0352}

\bibitem[{{Page} {et~al.}(2022){Page}, {Beardmore}, {Osborne}, {Munari},
  {Ness}, {Evans}, {Bode}, {Darnley}, {Drake}, {Kuin}, {O'Brien}, {Orio},
  {Shore}, {Starrfield}, \& {Woodward}}]{page22}
{Page}, K.~L., {Beardmore}, A.~P., {Osborne}, J.~P., {et~al.} 2022, \mnras,
  514, 1557, \dodoi{10.1093/mnras/stac1295}

\bibitem[{{Pavana} {et~al.}(2019){Pavana}, {Anche}, {Anupama}, {Ramaprakash},
  \& {Selvakumar}}]{pavana19}
{Pavana}, M., {Anche}, R.~M., {Anupama}, G.~C., {Ramaprakash}, A.~N., \&
  {Selvakumar}, G. 2019, \aap, 622, A126, \dodoi{10.1051/0004-6361/201833728}

\bibitem[{{Pavlenko} {et~al.}(2008){Pavlenko}, {Evans}, {Kerr}, {Yakovina},
  {Woodward}, {Lynch}, {Rudy}, {Pearson}, \& {Russell}}]{pavlenko08}
{Pavlenko}, Y.~V., {Evans}, A., {Kerr}, T., {et~al.} 2008, \aap, 485, 541,
  \dodoi{10.1051/0004-6361:20078622}

\bibitem[{{Pavlenko} {et~al.}(2016){Pavlenko}, {Kaminsky}, {Rushton}, {Evans},
  {Woodward}, {Helton}, {O'Brien}, {Jones}, \& {Elkin}}]{pavlenko16}
{Pavlenko}, Y.~V., {Kaminsky}, B., {Rushton}, M.~T., {et~al.} 2016, \mnras,
  456, 181, \dodoi{10.1093/mnras/stv2546}

\bibitem[{{Pizzuto} {et~al.}(2021){Pizzuto}, {Vandenbroucke}, {Santander}, \&
  {IceCube Collaboration}}]{pizzuto21}
{Pizzuto}, A., {Vandenbroucke}, J., {Santander}, M., \& {IceCube
  Collaboration}. 2021, The Astronomer's Telegram, 14851, 1

\bibitem[{{Razzaque} {et~al.}(2010){Razzaque}, {Jean}, \& {Mena}}]{soeb10}
{Razzaque}, S., {Jean}, P., \& {Mena}, O. 2010, \prd, 82, 123012,
  \dodoi{10.1103/PhysRevD.82.123012}

\bibitem[{{Shore} {et~al.}(2012){Shore}, {Wahlgren}, {Augusteijn}, {Liimets},
  {Koubsky}, {{\v{S}}lechta}, \& {Votruba}}]{shore12}
{Shore}, S.~N., {Wahlgren}, G.~M., {Augusteijn}, T., {et~al.} 2012, \aap, 540,
  A55, \dodoi{10.1051/0004-6361/201118060}

\bibitem[{{Shore} {et~al.}(2011){Shore}, {Wahlgren}, {Augusteijn}, {Liimets},
  {Page}, {Osborne}, {Beardmore}, {Koubsky}, {{\v{S}}lechta}, \&
  {Votruba}}]{shore11}
---. 2011, \aap, 527, A98, \dodoi{10.1051/0004-6361/201015901}

\bibitem[{{Sitarek} \& {Bednarek}(2012)}]{sitarek12}
{Sitarek}, J., \& {Bednarek}, W. 2012, \prd, 86, 063011,
  \dodoi{10.1103/PhysRevD.86.063011}

\bibitem[{{Sokoloski} {et~al.}(2006){Sokoloski}, {Luna}, {Mukai}, \&
  {Kenyon}}]{sokolski06}
{Sokoloski}, J.~L., {Luna}, G.~J.~M., {Mukai}, K., \& {Kenyon}, S.~J. 2006,
  \nat, 442, 276, \dodoi{10.1038/nature04893}

\bibitem[{{Sokolovsky} {et~al.}(2021){Sokolovsky}, {Aydi}, {Chomiuk}, {Kawash},
  {Strader}, {Babul}, {Sokoloski}, {Mioduszewski}, {Linford}, {Mukai}, {Li},
  {O'Brien}, \& {Rupen}}]{ATel14886}
{Sokolovsky}, K., {Aydi}, E., {Chomiuk}, L., {et~al.} 2021, The Astronomer's
  Telegram, 14886, 1

\bibitem[{Tatischeff \& Hernanz(2007)}]{tatischeff07}
Tatischeff, V., \& Hernanz, M. 2007, \apjl, 663, L101, \dodoi{10.1086/520049}

\bibitem[{{Tatischeff} \& {Hernanz}(2008)}]{tatischeff08}
{Tatischeff}, V., \& {Hernanz}, M. 2008, in Astronomical Society of the Pacific
  Conference Series, Vol. 401, RS Ophiuchi (2006) and the Recurrent Nova
  Phenomenon, ed. A.~{Evans}, M.~F. {Bode}, T.~J. {O'Brien}, \& M.~J.
  {Darnley}, 328, \dodoi{10.48550/arXiv.0801.1770}

\bibitem[{{Warner}(2003)}]{warner03}
{Warner}, B. 2003, {Cataclysmic Variable Stars},
  \dodoi{10.1017/CBO9780511586491}

\bibitem[{{Weiler} {et~al.}(2002){Weiler}, {Panagia}, {Montes}, \&
  {Sramek}}]{weiler02}
{Weiler}, K.~W., {Panagia}, N., {Montes}, M.~J., \& {Sramek}, R.~A. 2002,
  \araa, 40, 387, \dodoi{10.1146/annurev.astro.40.060401.093744}

\bibitem[{{Williams} {et~al.}(2021){Williams}, {O'Brien}, {Woudt}, {Nyamai},
  {Green}, {Titterington}, {Fender}, \& {Sivakoff}}]{ATel14849}
{Williams}, D., {O'Brien}, T., {Woudt}, P., {et~al.} 2021, The Astronomer's
  Telegram, 14849, 1

\bibitem[{{Zheng} {et~al.}(2022){Zheng}, {Huang}, {Zhang}, {Zhang}, {Liu}, \&
  {Wang}}]{zheng22}
{Zheng}, J.-H., {Huang}, Y.-Y., {Zhang}, Z.-L., {et~al.} 2022, \prd, 106,
  103011, \dodoi{10.1103/PhysRevD.106.103011}

\end{thebibliography}
\bibliographystyle{aasjournal}



\end{document}